\documentclass[twocolumn,preprintnumbers,amsmath,amssymb,superscriptaddress,nofootinbib,longbibliography]{revtex4-1}
\pdfoutput=1

\usepackage{graphicx}
\usepackage{bm}
\usepackage{dsfont}
\usepackage[usenames,dvipsnames]{xcolor}
\usepackage{pstricks}
\usepackage[tight]{subfigure}
\usepackage{verbatim}
\usepackage{units}
\usepackage{multirow}
\usepackage{enumitem}
\usepackage{mathrsfs}
\usepackage{leftidx}

\usepackage[a4paper]{hyperref}
\hypersetup{colorlinks=true,linktoc=all,linkcolor=blue,breaklinks=true,citecolor=blue,urlcolor=blue}

\usepackage[english]{babel}
\newcommand{\kFw}{k_\text{F}^\text{w}}
\newcommand{\vFw}{v_\text{F}^\text{w}}
\newcommand{\EFw}{E_\text{F}^\text{w}}

\newcommand{\kw}{k_{n}^\text{w}}
\newcommand{\Eg}{E_\text{g}}

\newcommand{\depsr}{\Delta\varepsilon_\text{r}}

\newcommand{\kq}{k_{\sigma}}
\newcommand{\kFq}{k_{\text{F}\sigma}}

\newcommand{\EF}{E_\text{F}}
\newcommand{\me}{m_\text{e}}

\newcommand{\via}{\emph{via} }

\begin{document}


\title{Gate-voltage response of a one-dimensional ballistic spin valve\\ without spin-orbit interaction}

\author{Maciej Misiorny}
\email{misiorny@amu.edu.pl}
\affiliation{Department of Microtechnology and Nanoscience MC2, Chalmers University of Technology, SE-412 96 G\"{o}teborg, Sweden}
\affiliation{Faculty of Physics, Adam Mickiewicz University, 61-614 Pozna\'{n}, Poland}
\author{Carola Meyer}
\affiliation{Fachbereich Physik, Universit\"at Osnabr\"uck, D-49069 Osnabr\"uck, Germany}
\affiliation{Peter Gr\"unberg Institut (PGI-6),  Forschungszentrum J\"ulich \& JARA--Fundamentals of Future Information Technologies - DE-52 425 J\"ulich, Germany}

\date{\today}

\begin{abstract}
We show that engineering of tunnel barriers forming at the interfaces of a one-dimensional spin valve provides a viable path to a strong gate-voltage tunability of the magnetoresistance effect. In particular, 	we investigate theoretically a carbon nanotube (CNT) spin valve in terms of the influence of the CNT-contact interface on the performance of the device. The focus is on the strength and the spin selectivity of the tunnel barriers that are modelled as Dirac-delta potentials. The scattering matrix approach is used to derive the transmission coefficient that yields the tunneling magnetoresistance (TMR). We find a strong non-trivial gate-voltage response of the TMR in the absence of spin-orbit coupling when the energy of the incident electrons matches the potential energy of the barrier. Analytic expressions for the TMR in various limiting cases are derived. These are used to explain previous experimental results, but also to predict prospective ways for device optimization with respect to size and tunability of the TMR effect in the ballistic transport regime by means of engineering the tunnel barriers at the CNT-contact interfaces.

\end{abstract}

\maketitle

\section{Introduction}

The already vast and still growing research area of spintronics offers a new functionality for solid-state devices based on the spin of electrons rather than their charge~\cite{Bader_Annu.Rev.Condens.MatterPhys.1/2010,Spintronics_InsightNatureMater11/2012}. In this context, graphene and carbon nanotubes (CNTs) are regarded as extremely promising materials~\cite{Hueso07,Dlubak12,Han14}, since the spin lifetime is long due to the small spin-orbit coupling and due to a low natural abundance of C$^{13}$ nuclear spins~\cite{guim14, Beschoten14, Laird13, Viennot15, Morgan16}. Additionally, the Fermi velocity of these materials is very high resulting in short dwell times within a device. This combination can, in turn,  lead to a large magnetoresistance (MR) effect as well as to a large absolute change of resistance ---both important for a good performance of spin-valve devices~\cite{Dlubak12}. A key issue for such devices is to enhance the spin injection efficiency by optimizing both the contact material~\cite{Hueso07} and the tunnel barrier~\cite{Dlubak12,Han10,Morgan16}.

Another major point, not discussed in depth so far, regards the fact that many interesting effects in spintronic devices stem from the spin-orbit coupling~\cite{Zutic04,Sinova_Rev.Mod.Phys.87/2015}. To tune the MR effect efficiently with a gate voltage, for instance, as in a spin transistor, a strong spin-orbit coupling (SOC) is desired. Though the SOC is very weak in graphene~\cite{Gmitra09}, it can be strongly enhanced by adatoms that induce local $sp^3$ hybridization in the carbon bonds~\cite{Castro09}. On the other hand, the spin-orbit interaction in CNTs is stronger for the same reason due to their curvature~\cite{Huertas06,Jhang10} and has been found to be even larger in some devices~\cite{Steele13}. A gate voltage will tune the MR effect in such materials rather efficiently, however, this is usually achieved on the expense of the spin relaxation time that is the great asset of carbon materials.

In this paper, we present model calculations that reveal another option for gate-controlled spin devices that avoids enhancing the spin-orbit coupling and therefore spin relaxation. 
We demonstrate that in quasi one-dimensional devices  the MR effect can show a strong tunability with gate voltage depending on the properties of tunnel barriers arising at the interfaces of the electrodes. In particular, we systematically analyze how the strength and spin-selectiveness of these tunnel barriers affect magneto-transport characteristics of a ballistic one-dimensional spin valve employing a CNT as model system. This aspect seems to be of key importance for full understanding of the injection of spin-polarized electrons into a CNT, and it has not been examined in full detail hitherto. 
Actually, although spin-dependent transport through a CNT attached to ferromagnetic metallic electrodes has been the subject of extensive experimental studies for almost two decades~\cite{Tsukagoshi_Nature401/1999,Kim_Phys.Rev.B66/2002,Zhao_Appl.Phys.Lett.80/2002,Jensen_Phys.Rev.B72/2005,Sahoo_Appl.Phys.Lett.86/2005,Man_Phys.Rev.B73/2006}, only recently the role of the tunnel barrier strength in this process has been addressed~\cite{Morgan16} ---showing that the MR of a device is significantly influenced by this factor.

For the purpose of this study, we consider a CNT-based spin valve that basically acts as an electronic interferometer, a setup employed formerly both in experiment~\cite{Liang_Nature411/2001,Sahoo_NaturePhys.1/2005,Man_Phys.Rev.Lett.95/2005} and theoretically~\cite{Cottet_Europhys.Lett.74/2006,Cottet_Sem.Sci.Tech.21/2006}. In order to capture the effect of electrode-CNT interfaces, at which tunnel barriers form, we treat them as spin-dependent Dirac-delta potentials. A similar approach has been used to study spin injection between a ferromagnetic metal and a two-dimensional electron system~\cite{Grundler_Phys.Rev.Lett.86/2001,Hu_Phys.Rev.Lett.87/2001}. Then, calculations of spin-dependent transport through a device are derived by means of the scattering matrix approach. We find that engineering of the strength and spin-selective properties of the tunnel barriers  in combination with the gate-voltage tuning provides a path for obtaining devices in which the magnitude of MR effect can be adjusted in a broad range from $-15.5$\% up to $+40$\%.

The paper is organized as follows; First, in Sec.~\ref{sec:Model} we introduce the model of a CNT-based spin valve, and define the concept of a tunnel barrier at the electrode-CNT interface. Next, in Sec.~\ref{sec:Tunneling_interface} we provide a theoretical description of spin injection through the interface and derive the corresponding transmission coefficient for conduction electrons. This enables us to determine the linear transport through the device as outlined in Sec.~\ref{sec:Transport_theory}. Numerical results are presented in Sec.~\ref{sec:Numerical_results}, where we discuss both the case of a single (Sec.~\ref{sec:Single_channel}) and many (Sec.~\ref{sec:Many_channels}) orbital transport channels. There, we consider in detail the limit of strong tunnel barriers (Sec.~\ref{sec:Strong_barriers}), as well as the situation when a device is characterized by the asymmetric (Sec.~\ref{sec:Asymm_bar}) and spin-selective (Sec.~\ref{sec:Spin_sel_bar}) barriers. Finally, we conclude the paper in Sec.~\ref{sec:Conclusions} with a discussion about possible implementations of such barriers regarding the essential effects and the general performance to be expected from prospective devices.

\section{\label{sec:Model}Model of a CNT-based spin valve}

\begin{figure}
	\includegraphics[width=0.875\columnwidth]{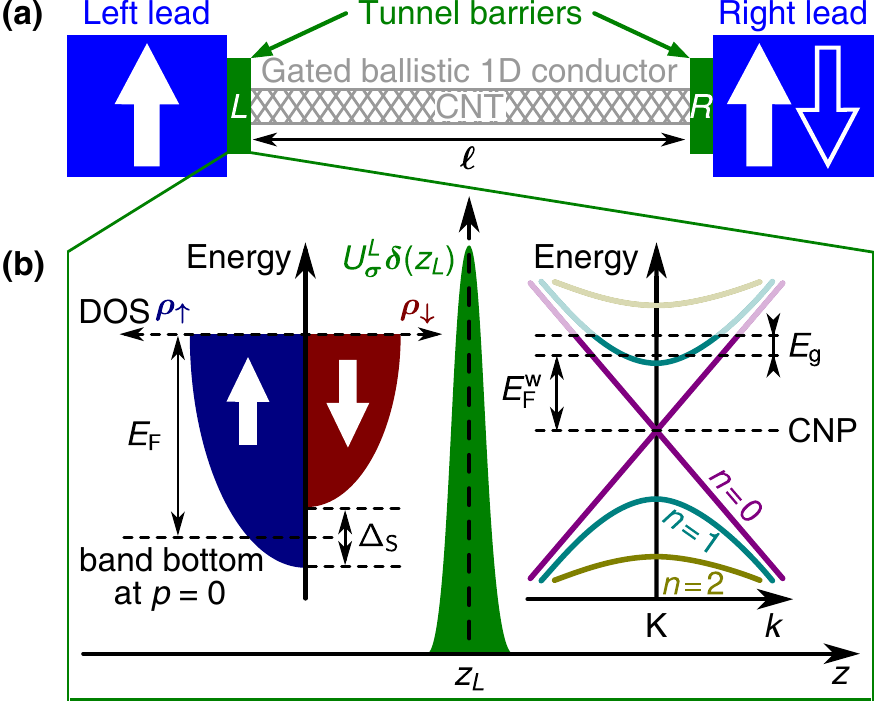}
	\caption{
		(a)~A graphic depiction of the model device: a gated CNT inserted between two ferromagnetic (FM) metallic leads, whose spin moments are oriented either parallel or antiparallel.
		(b)~Schematic representation of a tunnel barrier arising at the left CNT-lead interface that is modelled as a spin-selective repulsive Dirac-delta potential $U_\sigma^L\delta(z_L)$. 
		In the left (right) side of (b) an example dispersion relation for a FM lead (CNT around the Fermi point K) is shown. Here, CNP stands for the charge neutrality point with respect to which energy is measured.
		For a detailed description see Sec.~\ref{sec:Model}.
	}
	\label{fig1}
\end{figure}

A device under consideration consists of two ferromagnetic (FM) metallic leads interconnected by a CNT which we  approximate as a \emph{ballistic} and \emph{noninteracting} one-dimensional (1D) quantum wire~\cite{Kane_Phys.Rev.Lett.78/1997,Egger_Eur.Phys.J.B3/1998,Cottet_Sem.Sci.Tech.21/2006}, see Fig.~\ref{fig1}(a). Importantly, at both CNT-lead interfaces a tunnel barrier can form, whose exact shape, generally different for each interface, is unknown. For this reason, we model scattering of tunneling electrons at the interfaces by means of a \emph{spin-selective} repulsive Dirac-delta potential $U_\sigma^q\delta(z_q)$ for $q=L(\text{eft}),R(\text{ight})$, see Fig.~\ref{fig1}(b). Such an approach has already been shown to suffice in capturing key transport features of the interface~\cite{Blonder_Phys.Rev.B25/1982,Qi_Phys.Rev.B58/1998,Grundler_Phys.Rev.Lett.86/2001,Hu_Phys.Rev.Lett.87/2001}, but so far has not been systematically applied to analyze how its properties affect one-dimensional spin transport.

In the model to be analyzed, two identical FM leads are described as a reservoir of non-interacting, itinerant electrons within the Stoner model, with the dispersion relation given by
\begin{equation}
	\varepsilon_\sigma
	=
	\frac{\hbar^2(\kq)^2}{2m^\ast}
	-
	\eta_\sigma\frac{\Delta_\text{S}}{2}
	-
	\EF
	.
\end{equation}
Here, $\Delta_\text{S}$ denotes the Stoner splitting, $\eta_{\uparrow(\downarrow)}=\pm1$, and $\EF$ represents the Fermi energy ---note that energy $\varepsilon_\sigma$ is measured relative to the Fermi level. Additionally, we assume the effective mass to be equal to the electron's mass, $m^\ast\approx\me$. Generally, in a bulk system with a parabolic dispersion (i.e., for the free-electron model) the spin-dependent density of states (DOS) $\rho_\sigma$ at the Fermi level (per unit volume and per spin channel) is related to the spin-dependent Fermi wave vector, $\kFq\equiv\kq(\varepsilon_\sigma=0)$, as
$
	\rho_\sigma
	=
	2\me\kFq/(4\pi^2\hbar^2)
$,
as shown in the left side of Fig.~\ref{fig1}(b).
With this, we introduce the \emph{spin-polarization} coefficient $p$ for the material of which leads are made~\cite{Maekawa_book},
\begin{equation}
	p
	=
	\frac{
	\rho_+ - \rho_- 
	}{
	\rho_+ + \rho_-
	}
	,
\end{equation}
with the spin index $\eta=\pm$ referring now to \emph{spin-majori\-ty}~($+$) and \emph{spin-minority}~($-$) electrons.
Note that the notion of spin-majority/minority electrons becomes useful in the present case, because two different collinear configurations of spin moments of electrodes, that is, parallel~(P) and antiparallel~(AP),  will be considered. In particular, the orientation of a spin moment of the left electrode will be kept fixed, so that the relation between spin-`up'/-`down' electrons  and spin-majority/-minority electrons in the left electrode takes the following form
\begin{equation}\label{eq:def_spin_L}
	\text{P/AP}(\sigma_{L})
	\equiv
	\left\{
	\begin{aligned}
	+&\ \text{if}\ \sigma=\ \uparrow,
	\\
	-&\ \text{if}\ \sigma=\ \downarrow.
	\end{aligned}
	\right.
\end{equation}
This also sets the reference frame for spin orientations of electronic spins in the right electrode. As a result, when a spin moment of the right electrode is parallel/antiparallel with respect to the left one, we get, respectively,
\begin{equation}\label{eq:def_spin_R_P}
	\text{P}(\sigma_{R})
	\equiv
	\left\{
	\begin{aligned}
	+&\ \text{if}\ \sigma=\ \uparrow,
	\\
	-&\ \text{if}\ \sigma=\ \downarrow,
	\end{aligned}
	\right.
\end{equation}
and
\begin{equation}\label{eq:def_spin_R_AP}
	\text{AP}(\sigma_{R})
	\equiv
	\left\{
	\begin{aligned}
	+&\ \text{if}\ \sigma=\ \downarrow,
	\\
	-&\ \text{if}\ \sigma=\ \uparrow.
	\end{aligned}
	\right.	
\end{equation}
Moreover, in the limit of moderate spin polarizations observed in typical materials used for electrodes~\cite{Maekawa_book,Tsybmal_book} $\Delta_\text{S}/(2\EF)<1$, so that a wave vector can be approximated as $k_{\text{F}\eta}\approx k_0\big[1+\eta\Delta_\text{S}/(4\EF)\big]$ with $k_0=\sqrt{2\me\EF}/\hbar$ and, consequently, one can use the following parameterization of the Stoner splitting parameter $\Delta_\text{S}=4\EF p$. Note that the above approximation remains valid only for moderate values of the spin polarization of electrodes~($p<0.5$).

Next, essential features of a CNT in the vicinity of the Fermi point K (K$^\prime$) are captured by a dispersion relation~\cite{Kane_Phys.Rev.Lett.78/1997,Egger_Eur.Phys.J.B3/1998,DeMartino_J.Phys.:Condens.Matter17/2005},
\begin{equation}\label{eq:disp_rel_w}
	\varepsilon_{n}^\text{w}
	=
	\pm
	\hbar\vFw
	\sqrt{
		(\kw-\kFw)^2
		+
		(n/r)^2
	}
	+
	\Eg
	-
	\EFw
	,
\end{equation}
typical for 1D conductors~\cite{Giamarchi_book}, with the $+/-$ sign corresponding to conduction/valence band, see the right side of Fig.~\ref{fig1}(b). In Eq.~(\ref{eq:disp_rel_w}) $\vFw$ stands for the Fermi velocity and $n/r$ represents the quantized transverse momentum of a metallic CNT~\cite{DeMartino_J.Phys.:Condens.Matter17/2005}, with $r$ denoting the radius of a CNT and  $n=0,1,2,\ldots$ being the subband index. 
As above, the energy $\varepsilon_{n}^\text{w}$ is defined relative to the Fermi level~$\EFw$. Recall that for an undoped CNT it coincides with the charge neutrality point (CNP), i.e., $\EFw=0$, so that only one orbital channel ($n=0$) can contribute to transport at low temperature. However, due to modification of the immediate environment of a CNT $\EFw$ can be shifted by as much as $\pm1$~eV~\cite{Krueger_App.Phys.Lett.78/2001,Krueger_NewJ.Phys.5/2003}, and, thus, more channels become available for transport. The Fermi level can be further adjusted by application of an external gate voltage which leads to the shift $\Eg$ due to the capacitive coupling between  the gate and a CNT~\cite{Krueger_App.Phys.Lett.78/2001}. Note that Eq.~(\ref{eq:disp_rel_w}) remains valid as long as the variation in $\Eg$ is small, that is, the Fermi level is moderately shifted around $\EFw$.  It is assumed that transport of electrons along a CNT is ballistic and no mixing of channels occurs.

Finally, before we turn to the discussion of electron tunneling through the electrode-CNT interface, we would like to briefly comment on applicability limits of the model under consideration. We recall that electrodes are here approximated by only free ($s$-band) electrons, and tunnel barriers are treated as a Dirac-delta potential. In fact, the tunnel barrier forming at the electrode-CNT interface can be of much more complex nature, with a potential profile determined by additional factors not included in the present considerations, like the interface roughness and adsorbates~\cite{Heinze02,Foa-Torres_book}. Furthermore,  materials typically used for electrodes involve transition metals and their alloys, in case of which the free-electron model may be insufficient to capture all key features. In particular, for these materials a more complicated band structure is expected to underlie tunneling of electrons across the interface~\cite{Zhang_Eur.Phys.J.B10/1999}. In order to accommodate fully all these intricacies, that is, the complex electrode-CNT hybridization and the exact morphology of the interface, a model from first principles is needed~\cite{Mavropoulos_Phys.Rev.B69/2004,Nemec_Phys.Rev.Lett.96/2006}. Nevertheless, the present approach already shows the great potential of one-dimensional CNT spin valves in the ballistic transport regime with respect to size and tunability of the MR effect. 


\section{\label{sec:Tunneling_interface}Tunneling through a FM-metal/CNT interface}

Spin injection across an interface with the band structure mismatch at the Fermi energy has already been addressed, e.g., for FM-metal/metal~\cite{Gijs_Adv.Phys.46/1997} and FM-metal/semiconductor heterojunctions~\cite{Grundler_Phys.Rev.Lett.86/2001,Hu_Phys.Rev.Lett.87/2001}. Here, we consider a spin-dependent tunneling of electrons through the FM-metal/CNT interface, as illustrated in Fig.~\ref{fig1}(b). The relevant transmission coefficient $\mathcal{T}$  can be derived by means of standard quantum mechanical methods. 
The key problem  one has to face is then how to match the wave functions at the interface.
Let us focus on the left interface for the moment.

For an ideal interface (i.e., without spin-flip and inelastic/interchannel scattering)  the particle current $j_{\sigma n}^z$ along the $z$ axis across the interface has to be conserved in each spin ($\sigma$) and orbital ($n$) channel. This basically means that the current in the vicinity of the barrier on its left side, $j_{\sigma n}^{z}(z_L^-)$, has to match that on the right side, $j_{\sigma n}^{z}(z_L^+)$, namely,
$
	j_{\sigma n}^{z}(z_L^-)
	=
	j_{\sigma n}^{z}(z_L^+)
$
with $z_L^\pm\equiv z_L\pm0^+$ and $0^+$ denoting an infinitesimally small displacement. 
Close to the interface, on its \emph{left} side  (\mbox{$z<z_L$}),  corresponding to a FM metal, this current is given by
\begin{multline}
	j_{\sigma n}^{z}(z<z_L)
	=
	\frac{i\hbar}{2\me}
	\big[
	\big(\partial_z \bm{\Psi}_{\sigma n}^{\dagger}(z)\big)
	\underline{\bm{\sigma}}^0
	\bm{\Psi}_{\sigma n}(z)
	\\
	-
	\bm{\Psi}_{\sigma n}^{\dagger}(z)
	\underline{\bm{\sigma}}^0
	\big(\partial_z \bm{\Psi}_{\sigma n}(z)\big)
	\big]
	,
\end{multline}
whereas on the \emph{right} side (\mbox{$z>z_L$}), that is, in a CNT, it takes the form
\begin{equation}
	j_{\sigma n}^{z}(z>z_L)
	=
	\vFw
	\bm{\Phi}_{\sigma n}^{\dagger}(z)
	\underline{\bm{\sigma}}^z
	\bm{\Phi}_{\sigma n}(z)
	,
\end{equation}
with $\underline{\bm{\sigma}}^z$ ($\underline{\bm{\sigma}}^0$) denoting the Pauli (identity) matrix, and the wave functions $\bm{\Psi}$ and $\bm{\Phi}$ defined as
\begin{equation}\label{eq:wave_fun_Psi}
	\bm{\Psi}_{\sigma n}(z)
	=
	\begin{pmatrix}
	\psi_{\sigma n}^{\rightarrow}
	\text{e}^{ik_\sigma z}
	\\
	\psi_{\sigma n}^{\leftarrow}
	\text{e}^{-ik_\sigma z}
	\end{pmatrix}
	,
\end{equation}
and
\begin{equation}\label{eq:wave_fun_Phi}
	\bm{\Phi}_{\sigma n}(z)
	\!
	=
	\!
	\begin{pmatrix}
	\phi_{\sigma n}^{\rightarrow}
	\text{e}^{i\kw z}
	\\
	\phi_{\sigma n}^{\leftarrow}
	\text{e}^{-i\kw z}
	\end{pmatrix}
	.
\end{equation}
Here, $\psi_{\sigma n}^{d}$ and $\phi_{\sigma n}^{d}$ generally represent the respective probability amplitude for right (\mbox{$d=\, \rightarrow$}) and left (\mbox{$d=\, \leftarrow$}) moving electrons. Inserting Eqs.~(\ref{eq:wave_fun_Psi})-(\ref{eq:wave_fun_Phi}) into the expressions for $j_{\sigma n}^{z}(z)$, one obtains
$
	j_{\sigma n}^{z}(\mbox{$z<z_L$})	
	=
	(\hbar \kq/\me)
	\big[|\psi_{\sigma n}^{\rightarrow}|^2-|\psi_{\sigma n}^{\leftarrow}|^2\big]
$
and 
$
	j_{\sigma n}^{z}(\mbox{$z>z_L$})	
	=
	\vFw
	\big[|\phi_{\sigma n}^{\rightarrow}|^2-|\phi_{\sigma n}^{\leftarrow}|^2\big]
$.
In consequence, one can define the transmission amplitude for electrons incident on the left  interface from left (`$\rightarrow$') / right (`$\leftarrow$') in terms of flux amplitudes 
as 
\begin{equation}
	\mathcal{T}_{\sigma n}^{L\rightarrow}
	=
	\Bigg|
	\frac{
	\sqrt{\vFw}
	\phi_{\sigma n}^{\rightarrow}
	}{
	\sqrt{\hbar \kq/\me}
	\psi_{\sigma n}^{\rightarrow}
	}
	\Bigg|^2
\end{equation}
and
\begin{equation}
	\mathcal{T}_{\sigma n}^{L\leftarrow}
	=
	\Bigg|
	\frac{
	\sqrt{\hbar \kq/\me}
	\psi_{\sigma n}^{\leftarrow}
	}{
	\sqrt{\vFw}
	\phi_{\sigma n}^{\leftarrow}
	}
	\Bigg|^2
	.
\end{equation}
Analogous definitions also hold for the right interface.
Interestingly, one can note that the same result for $j_{\sigma n}^{z}(\mbox{$z>z_L$})$ can be reached 
if one used the free-electron model, for which
\begin{multline}
	j_{\sigma n}^{z}(z>z_L)
	=
	\frac{i\hbar}{2m_n^\ast}
	\big[
	\big(\partial_z \bm{\Phi}_{\sigma n}^{\dagger}(z)\big)
	\underline{\bm{\sigma}}^0
	\bm{\Phi}_{\sigma n}(z)
	\\
	-
	\bm{\Phi}_{\sigma n}^{\dagger}(z)
	\underline{\bm{\sigma}}^0
	\big(\partial_z \bm{\Phi}_{\sigma n}(z)\big)
	\big]
	,
\end{multline}
with the effective mass $m_n^\ast=\hbar\kw/\vFw$. 
For this reason, the continuity of the current across the $q$th interface between a FM lead and a CNT can be ensured by imposing the following boundary conditions for wave functions~\cite{Kroemer_J.Vac.Sci.Technol.21/1982,Zhu_Phys.Rev.B27/1983,Harrison_J.Appl.Phys.110/2011}:
\begin{equation}
	\sqrt{\frac{M_n}{\me}}
	\text{Tr}\big[
	\underline{\bm{\sigma}}^0
	\bm{\Psi}_{\sigma n}(z_q)
	\big]
	=
	\sqrt{\frac{M_n}{m_n^\ast}}
	\text{Tr}\big[
	\underline{\bm{\sigma}}^0
	\bm{\Phi}_{\sigma n}(z_q)
	\big]
	,
\end{equation}
\begin{multline}
	\sqrt{\frac{M_n}{m_n^\ast}}
	\text{Tr}\big[
	\underline{\bm{\sigma}}^0
	\partial_z
	\bm{\Phi}_{\sigma n}(z)|_{z_q}
	\big]
	-
	\sqrt{\frac{M_n}{\me}}
	\text{Tr}\big[
	\underline{\bm{\sigma}}^0
	\partial_z
	\bm{\Psi}_{\sigma n}(z)|_{z_q}
	\big]
	\\
	=
	\frac{2M_nU_\sigma^q}{\hbar^2}
	\sqrt{\frac{M_n}{m_n^\ast}}
	\text{Tr}\big[
	\underline{\bm{\sigma}}^0
	\bm{\Phi}_{\sigma n}(z_q)
	\big]
	,
\end{multline}
with $M_n=\sqrt{\me m_n^\ast}$. 
The transmission coefficient $\mathcal{T}_{\sigma n}^q\equiv\mathcal{T}_{\sigma n}^{q\rightarrow}=\mathcal{T}_{\sigma n}^{q\leftarrow}$, which characterizes tunneling of an electron with spin $\sigma$ to/out the $n$th channel of a CNT across the $q$th interface,  takes thus the following form ($\varepsilon=\varepsilon_\sigma=\varepsilon_n^\text{w}$)
\begin{equation}\label{eq:Tq}
	\mathcal{T}_{\sigma n}^{qc}(\varepsilon)
	=
	\frac{
		4k_{c(\sigma_q)}(\varepsilon)\kw(\varepsilon)	
	}{
	\big[k_{c(\sigma_q)}(\varepsilon)+\kw(\varepsilon)\big]^2
	+
	\big[Z_\sigma^q\big]^2
	\kw(\varepsilon)\kappa
	}
	.
\end{equation}
The action of the magnetic configuration index~$c(\sigma_q)$, which for a given configuration $c=\text{P,AP}$ relates  spin-$\sigma_q$ electrons to spin-majority/-minority electrons in the $q$th electrode, should be interpreted  by means of  Eqs.~(\ref{eq:def_spin_L})-(\ref{eq:def_spin_R_AP}).
Furthermore, $\kappa=2\EF/(\hbar\vFw)$, and $Z_\sigma^q=k_0U_\sigma^q/\EF$ is the spin-selective \emph{dimensionless barrier strength}, defined as the ratio of the spin-dependent potential energy of the barrier $k_0U_\sigma^q$ and the energy of an incident electron from the Fermi level of a lead. 
Here, we additionally introduce the spin asymmetry parameter $\alpha_q$ for the $q$th barrier, 
\begin{equation}\label{eq:alpha}
	\alpha_q
	=
	\frac{
	Z_\uparrow^q-Z_\downarrow^q
	}{
	Z_\uparrow^q+Z_\downarrow^q
	}
	,
\end{equation}
so that 
$
	Z_\sigma^q
	=
	Z_q(1+\eta_\sigma\alpha_q)
$
and $Z_q=(Z_\uparrow^q+Z_\downarrow^q)/2$ with $-1<\alpha_q<1$. Note that a positive (negative) $\alpha_q$ means that the probability for spin-down (spin-up) electrons to tunnel through the barrier is higher due to a smaller barrier strength. The limit of $\alpha_q\rightarrow+1\,(-1)$ corresponds then to a  vanishingly small barrier, i.e., almost perfect transmission, for spin-down (spin-up) electrons.
Importantly, the spin selectiveness of a tunnel barrier, characterized by the parameter~$\alpha_q$, is an inherent property of the barrier and it is not associated with the magnetic configuration of electrodes. In particular, note that in Eq.~(\ref{eq:Tq}) the magnetic configuration index $c(\sigma_q)$ affects only wave vectors of electrons in the electrode.
This effect should not be confused with the spin dependence of the transmission coefficient $\mathcal{T}_{\sigma n}^{qc}(\varepsilon)$ of a barrier, which involves both effects of the barrier spin selectiveness (determined by the spin asymmetry $\alpha_q$) and magnetic properties of electrodes (characterized  both by the spin polarization $p$ and the magnetic configuration of the valve).

\section{\label{sec:Transport_theory}Linear transport through a CNT-based spin valve}

Within the scattering matrix approach, the linear response conductance at temperature $T$ is given by~\cite{Datta_book} 
\begin{equation}\label{eq:G_def}
\hspace*{-3pt}
	G_{\text{P}\!/\!\text{AP}}
	=
	\frac{e^2}{h}
	\!
	\cdot
	\!
	\frac{1}{4k_\text{B}T}
	\!
	\sum_{n\sigma}
	\!
	\int\!\!\text{d}\varepsilon\,
	\mathscr{T}_{\sigma n}^{\text{P}\!/\!\text{AP}}(\varepsilon)
	\cosh^{-2}
	\!
	\Big(
	\frac{\varepsilon}{k_\text{B}T}
	\Big)
	,
	\!
\end{equation}
where  $\mathscr{T}_{n\sigma}^{P/AP}\!(\varepsilon)$ stands for the transmission coefficient of an electron with spin $\sigma$ passing through the $n$th channel of a device in the P/AP magnetic configuration~\cite{Datta_book,Cottet_Europhys.Lett.74/2006}, 
\begin{equation}\label{eq:TT_def}
\hspace*{-2pt}
	\mathscr{T}_{\sigma n}^{c}(\varepsilon)
	=
	\frac{
		\mathcal{T}_{\sigma n}^{Lc}(\varepsilon)\mathcal{T}_{\sigma n}^{Rc}(\varepsilon)
	}{
	\Big|
	1\!-\!\sqrt{
		\big(1\!-\!\mathcal{T}_{\sigma n}^{Lc}(\varepsilon)\big)
		\big(1\!-\!\mathcal{T}_{\sigma n}^{Rc}(\varepsilon)\big)
	}
	\,
	\text{e}^{i\theta_{\sigma n}^{c}\!(\varepsilon)}
	\Big|^2
	}
	.
	\!
\end{equation}
In the equation above, 
$
	\theta_{\sigma n}^{c}(\varepsilon)
	=
	2\delta_n(\varepsilon)+\varphi_{\sigma n}^{Lc}(\varepsilon)+\varphi_{\sigma n}^{Rc}(\varepsilon)
$
is the quantum-mechanical phase an electron acquires during its resonant transport through a CNT.
Here, the first term  
$
	\delta_n(\varepsilon)
	=
	\ell\big[
	\kFw
	+
	\sqrt{(\varepsilon+\EFw-\Eg)^2/(\hbar\vFw)^2-(n/r)^2}
	\big]
$,
cf. Eq.~(\ref{eq:disp_rel_w}), corresponds to the phase stemming from the ballistic propagation of an electron between the opposite interfaces of a CNT of the length $\ell$, while the second one $\varphi_{\sigma n}^{Lc}(\varepsilon)+\varphi_{\sigma n}^{Rc}(\varepsilon)$ represents the spin-dependent interfacial phase shift~\cite{Cottet_Europhys.Lett.74/2006} that arises when an electron is scattered at the left ($L$) and right ($R$) interface back into a CNT. This shift is basically related to the reflection amplitudes as $\varphi_{\sigma n}^L(\varepsilon)=\arg\big(r_{\sigma n}^{L\leftarrow}(\varepsilon)\big)$ and $\varphi_{\sigma n}^R(\varepsilon)=\arg\big(r_{\sigma n}^{R\rightarrow}(\varepsilon)\big)$, with the amplitudes at the interfaces defined as $r_{\sigma n}^{L\leftarrow}=\phi_{\sigma n}^{\rightarrow}/\phi_{\sigma n}^{\leftarrow}$ and $r_{\sigma n}^{R\rightarrow}=\phi_{\sigma n}^{\leftarrow}/\phi_{\sigma n}^{\rightarrow}$, so that one finds
\begin{equation}\label{eq:SDIPS}
\hspace*{-4pt}
	\varphi_{\sigma n}^{qc}(\varepsilon)
	=
	\arg\!\Bigg(
	\!
	\frac{
	-k_{c(\sigma_q)}(\varepsilon)+\kw(\varepsilon)-iZ_\sigma^q\sqrt{\kw(\varepsilon)\kappa}	
	}{
	k_{c(\sigma_q)}(\varepsilon)+\kw(\varepsilon)+iZ_\sigma^q\sqrt{\kw(\varepsilon)\kappa}
	}
	\Bigg)
	\!
	.
	\!\!
\end{equation}

Finally, one can note that in the limit of low temperature only electrons from the vicinity of the Fermi level ($\varepsilon=0$), that is, from the energy window of a few $k_\text{B}T$ around the Fermi level, contribute to transport, cp. Eq.~(\ref{eq:G_def}). At such an energy scale wave vectors $k_\sigma(\varepsilon)$ and $\kw(\varepsilon)$ vary insignificantly, and in consequence the change of $\mathcal{T}_{\sigma n}^q(\varepsilon)$ and $\varphi_{\sigma n}^q(\varepsilon)$ with energy is negligibly small. Therefore in the following discussion we assume $\mathcal{T}_{\sigma n}^{qc}(\varepsilon)\approx\mathcal{T}_{\sigma n}^{qc}(0)\equiv\mathcal{T}_{\sigma n}^{qc}$ and $\varphi_{\sigma n}^{qc}(\varepsilon)\approx\varphi_{\sigma n}^{qc}(0)\equiv\varphi_{\sigma n}^{qc}$.

\section{\label{sec:Numerical_results}Numerical results and discussion}

In order to discuss the dependence of spin-dependent transport through a CNT-based spin valve on the strength and properties of tunneling barriers at the interfaces, we consider a model CNT of a length $\ell=100$~nm and radius $r=2$~nm, characterized by $\vFw=8\times10^6$~m/s and $\kFw=8.5$~nm$^{-1}$~\cite{Liang_Nature411/2001}. 
As a result, the spacing between the subbands at the Fermi point amounts to $\Delta E\approx260$~meV. 
Furthermore, we assume that electrodes are described by the Fermi energy $\EF=8.5$~eV and the spin polarization parameter $p=0.25$. Such a value of~$p$ is very realistic, since common contact materials for CNTs, such as Permalloy and CoPd, exhibit this degree of spin-polarized injection of electrons~\cite{Moodera99, Morgan16}. 

The change of transport properties of a spintronic device when switching between the parallel and antiparallel magnetic configuration is generally captured by the tunneling magnetoresistance (TMR)
\begin{equation}
	\text{TMR}
	=
	\frac{G_\text{P}-G_\text{AP}}{G_\text{AP}}
	.
\end{equation}
If TMR is positive (negative), this basically means that conductance of the device is higher in the parallel (antiparallel) magnetic configuration than in the antiparallel (parallel) one.

In order to gain better insight into expected effects, first, in Sec.~\ref{sec:Single_channel}, we will consider a conceptually simplest case, that is, with only one orbital channel ($n=0$) available for transport. Such a case remains physically valid as long as the Fermi level of a CNT lies in the vicinity of the charge neutrality point, $|\EFw|\ll\Delta E$, so that at low temperatures the contribution of orbital channels (subbands) with $n\neq0$ to transport can be neglected, see the right side of Fig.~\ref{fig1}(b).  Later on, in Sec.~\ref{sec:Many_channels}, we will abandon this constraint and also discuss the case of many orbital channels by assuming that the Fermi level is shifted away from the charge neutrality point.

\subsection{\label{sec:Single_channel}The case of a single orbital channel}

\begin{figure}[t]
	\includegraphics[scale=1]{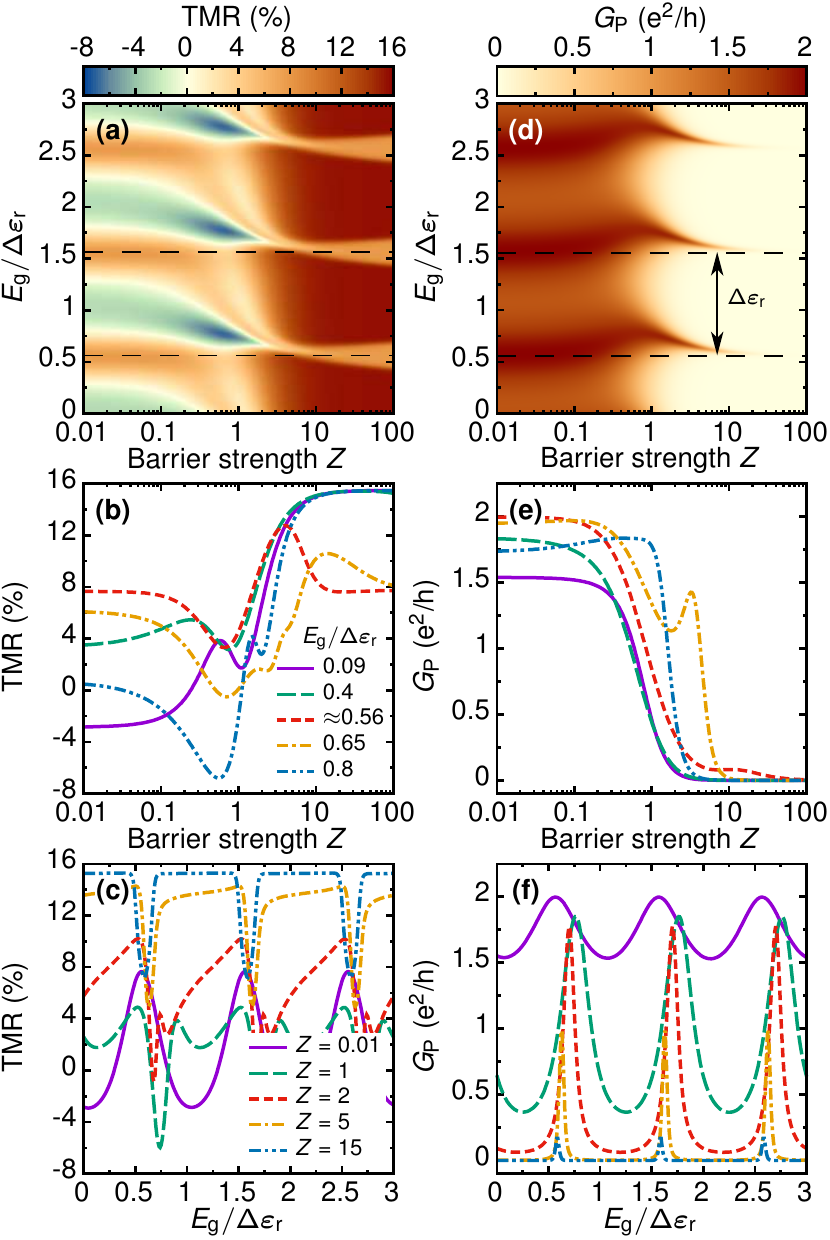}
	\caption{
		(color online) Density maps of tunneling magnetoresistance TMR (a) and conductance $G_\text{P}$ in the parallel magnetic configuration (d) shown as functions of the barrier strngth $Z$ and the shift of the Fermi level due to a gate voltage~$\Eg$ in the case of two \emph{identical} tunnel barriers at $T=2$~K. Here, $\depsr=\pi\hbar\vFw/\ell\ (\approx16.5\ \text{meV})$ stands for the period of oscillations.
		Horizontal thin dashed lines indicate two example values of $\Eg/\depsr\approx0.56$ and 1.56 at which resonant dips in TMR occur for large $Z$. 
		The middle panels (b) and (e) display the cross-sections of (a) and (d), respectively, for selected values of $\Eg/\depsr$ [see the legend in (b)], whereas the bottom panels (c) and (f) are analogous cross-sections but now for chosen values of $Z$ [see the legend in (c)].
		Other parameters as specified in the main text. 
	}
	\label{fig2}
\end{figure}

The hallmark of the model under discussion is the presence of the interference pattern in transport characteristics, as one can see in Fig.~\ref{fig2} where TMR and conductance are plotted for a device with two \emph{identical} tunnel barriers ($Z_\sigma^L=Z_\sigma^R=Z$). It is clear that such a  pattern in TMR stems directly from the periodic behavior of conductance as a function of the shift of the Fermi level due to a gate voltage $\Eg$, see Figs.~\ref{fig2}(d) and~\ref{fig2}(f). Since the conductance of the device $G$, Eq.~(\ref{eq:G_def}), is essentially determined by its transmission coefficient $\mathscr{T}$, Eq.~(\ref{eq:TT_def}), one can analyze $\mathscr{T}$ to obtain some basic information about the nature of such oscillations.

\begin{figure}[t]
	\includegraphics[scale=1]{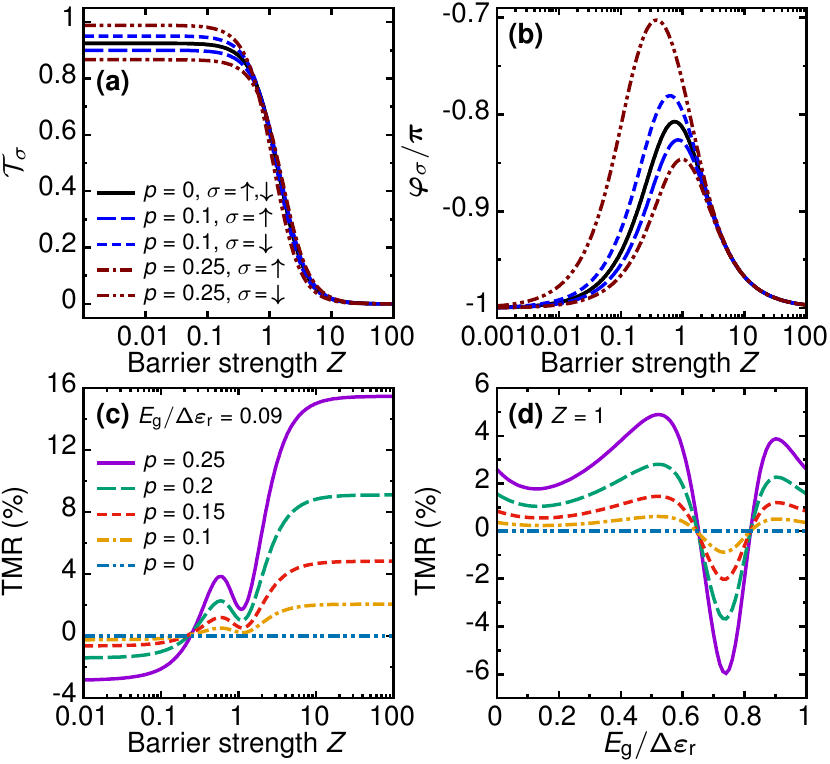}
	\caption{
		(color online)
		The effect of spin-polarized electrodes, quantified by spin-polarization coefficient $p$, on transport properties of a CNT-based spin valve with two identical tunnel  barriers. 
		\emph{Top panel} [(a)-(b)]: spin-dependent transmission coefficient~$\mathcal{T}_\sigma$ (a) and interfacial phase shift~$\varphi_\sigma$~(b) of a single barrier shown as functions of the barrier strength~$Z$. Here,  $\sigma=\uparrow(\downarrow)$ refers to spin-majority (-minority) electrons. 
		 {For description of lines see the legend in (a).}  
		\emph{Bottom panel} [(c)-(d)]: (c) dependence of tunneling magnetoresistance TMR on the barrier strength $Z$ for  $\Eg/\depsr\approx0.09$ ---the solid line is identical with the solid line in Fig.~\ref{fig2}(b); (d) TMR plotted as a function of $\Eg/\depsr$ for $Z=1$ ---the solid line is identical with the long-dashed line in Fig.~\ref{fig2}(c). 
		 {Corresponding lines in (c) and~(d) represent the same value of~$p$ as given in~(c).}
		Remaining parameters as in Fig.~\ref{fig2}. 
	}
	\label{fig3}
\end{figure}

From Eq.~(\ref{eq:TT_def}) one immediately finds that for a given $Z$ the transmission coefficient reaches its maximal achievable value at resonant energies for $p\in\mathbb{Z}$
\begin{equation}\label{eq:res_position}
	\widetilde{\varepsilon}^c_{\sigma p}
	=
	\depsr
	\Big[
	p
	-
	\dfrac{1}{2\pi}
	\big(
	\varphi_\sigma^{Lc}
	+
	\varphi_\sigma^{Rc}
	\big)
	\Big]
	-
	\hbar\vFw\kFw
	+
	\Eg
	,
\end{equation}
with $\depsr=\pi\hbar\vFw/\ell$ denoting the distance between consecutive resonances. Recall that here $\EFw=0$, which basically means that only one orbital channel ($n=0$) is active in transport. Thus, for the sake of notational clarity, in the remaining part of the present section we omit the orbital channel (subband) index~$n$. Importantly, one should notice that the position of these resonant states with respect to the Fermi level can be adjusted by application of a gate voltage, contributing \via $\Eg$. As a result,  whenever $\widetilde{\varepsilon}^c_{\sigma p}=0$ one observes resonant tunneling of electron through a device, which manifests as increased conductance, as shown in Figs.~\ref{fig2}(d) and~\ref{fig2}(f). Moreover, it should be emphasized that $\widetilde{\varepsilon}^c_{\sigma p}$ depends indirectly also on the strength of tunnel barriers $Z$ \via the spin-dependent interfacial phase shifts $\varphi_\sigma^{q}$ [see Fig.~\ref{fig3}(b)], and, consequently, also on the magnetic configuration of electrodes. This effect is especially observable in the nontrivial behavior of the TMR for small values of $Z$, see, e.g., the long-dashed line for $Z=1$ in Fig.~\ref{fig2}(c) where the energy of an incident electron matches the potential energy of the barrier. In the opposite limit of large $Z$, on the other hand, only sharp resonant dips in TMR can be observed, see the double-dotted-dashed line for $Z=15$ in Fig.~\ref{fig2}(c). Note that for small barriers $Z=1$ the TMR can be tuned between $-8\%$ and $+4\%$, see the long-dashed line in Fig.~\ref{fig2}(c). These are rather large values considering the high conductance in this regime compared to the results in Man \emph{et al.}~\cite{Man_Phys.Rev.B73/2006}. It is therefore important, while fabricating devices, to keep in mind that the length and the barrier strength will affect the tuning of the TMR effect with gate voltage. In general, it can be seen that the maxima in conductance, and consequently also in a TMR signal, arise owing to the phase factor~$\theta_\sigma^c(\varepsilon)$ occurring in the transmission coefficient~(\ref{eq:TT_def}). It is, thus, essential to keep track of this phase when simulating experimental data, and the present approach, which straightforwardly relates both the interface transmission~(\ref{eq:Tq}) and the interfacial phase shift~(\ref{eq:SDIPS}) to the strength of a tunnel barrier forming at the interface, proves to be useful to do it consistently.

To understand how the strength of tunnel barriers affects TMR, as shown in Figs.~\ref{fig2}(a) and~\ref{fig2}(b), let us analyze the dependence of spin-dependent transmission $\mathcal{T}_\sigma$ and interfacial phase shift $\varphi_\sigma$ of a single tunnel barrier, see Figs.~\ref{fig3}(a) and~\ref{fig3}(b). 
First of all, in Fig.~\ref{fig2}(a) one can distinguish three generic regions with respect to the barrier strength $Z$:  for \emph{small} $Z\lesssim0.1$ and \emph{large} $Z\gtrsim10$ where TMR remains roughly constant, and a \emph{transitional region} ($0.1\lesssim Z\lesssim10$) where TMR changes significantly. Interestingly, the occurrence of these can be explained by considering the behavior of $\mathcal{T}_\sigma$ and $\varphi_\sigma$ as a function~of~$Z$.

For small $Z$, a single barrier is characterized by a high transmission coefficient, with $\mathcal{T}_\uparrow\neq\mathcal{T}_\downarrow$ if $p\neq0$, Fig.~\ref{fig3}(a), and the interfacial phase shifts being close to~$-\pi$, Fig.~\ref{fig3}(b).  As a result, resonances in TMR, which originate from $\mathcal{T}_\uparrow\neq\mathcal{T}_\downarrow$, become only weakly shifted with respect to $\Eg$ as $Z$ is increased, see Eq.~(\ref{eq:res_position}). Note that even in the absence of spin polarization ($p=0$) the interface does not become fully transparent, that is, the transmission coefficient is still less than 1, see the solid line in Fig.~\ref{fig3}(a). This stems from the electronic-band structure mismatch between a lead and a CNT, which effectively manifests as different wave vectors for the lead,~$k_{c(\sigma_q)}$, and the CNT,~$\kw$, in Eq.~(\ref{eq:Tq}).
Further increase of $Z$ into the transitional region leads to a rapid drop of $\mathcal{T}_\sigma$, and to an increase of $\varphi_\sigma$. The maximum of $\varphi_\sigma$ shifts with $Z$ depending on $p$ and $\sigma$, Fig.~\ref{fig3}(b). Noteworthily, in this region a significant difference between $\varphi_\uparrow$ and $\varphi_\downarrow$ develops, which, in turn, means that resonant energies $\widetilde{\varepsilon}^c_{\sigma p}$ get markedly different for different spin orientations and magnetic configurations. This, in combination with the fact that in the $Z$-range under consideration a transition from  $\mathcal{T}_\uparrow<\mathcal{T}_\downarrow$ to $\mathcal{T}_\uparrow>\mathcal{T}_\downarrow$ occurs, leads to great changes in TMR preceded with a large shift of the resonances with respect to $\Eg$.
Finally, for large $Z$ the barriers become almost non-transparent, with the interfacial phase shift  approaching again $-\pi$ and $\varphi_\uparrow=\varphi_\downarrow$. Consequently, for asymptotically large $Z$ one observes constant TMR with narrow resonant dips appearing at exactly the same values of $\Eg$ as the resonant peaks in the limit of $Z\rightarrow0$.

To conclude the present discussion, in Figs.~\ref{fig3}(c)-(d) we additionally show how the main features of TMR as function of $Z$ considered above depend on the spin polarization of electrodes. The TMR effect is increasing with the polarization of the contacts for strong barriers just as in conventional spin valves, see Fig.~\ref{fig3}(c). Interestingly, the non-trivial behavior of the TMR around $Z=1$ is also more pronounced for larger polarization and, thus, the tunability of the TMR with the gate voltage as shown in Fig.~\ref{fig3}(d).

Finally, we would like to comment on the behavior of TMR in the limit of $Z\rightarrow0$. 
In general, one expects that in the experimental situation of electrical spin (diffusive) injection from a ferromagnet into a nonmagnetic material, the spin polarization of injected current can be quenched due to the conductance mismatch of these two materials ---the effect especially pronounced if the spin injection occurs into a semiconductor (SC)~\cite{Schmidt_Phys.Rev.B62/2000,Schmidt_J.Phys.D:Appl.Physi.38/2005}. Moreover, the conductance mismatch then essentially means that the transmission coefficient becomes spin-independent. This problem can be, however, circumvented if a spin-dependent interface resistance (e.g., due to a tunnel barrier), with some threshold value related to the resistivity and spin diffusion length of a SC, is introduced~\cite{Rashba_Phys.Rev.B62/2000,Fert_Phys.Rev.B64/2001}.
In the present considerations, on the other hand, such an effect is not captured by the model under investigation, that is, a CNT treated as a ballistic 1D conductor. Here, it is assumed that once an electron tunneled into the CNT its spin remains coherent until it tunnels out, which basically corresponds to the situation of both the spin diffusion length and the mean free path being sufficiently long. As shown by Valet and Fert~\cite{Valet_Phys.Rev.B48/1993}, in such a ballistic limit the usage of the Landauer approach is justified, without the need of applying 
the description of spin-dependent electrochemical potentials by means of the diffusion equation. 
Importantly, for that reason, in the current case the spin-dependence is preserved also in the limit of vanishingly small tunnel barrier, and consequently, a non-zero TMR signal is obtained.
We note that a similar effect was also derived for a spin injection into a SC in a ballistic picture~\cite{Grundler_Phys.Rev.Lett.86/2001,Hu_Phys.Rev.Lett.87/2001}.

\subsubsection{\label{sec:Strong_barriers}Limit of strong tunnel barriers}

In order to develop the complete physical picture, let us now briefly analyze transport in the case of large $Z$, which has been already widely studied~\cite{Liang_Nature411/2001,Cottet_Europhys.Lett.74/2006,Cottet_Sem.Sci.Tech.21/2006}. To begin with, in such a limit one generally derives
\begin{equation}\label{eq:T_approx}
	\mathcal{T}_\sigma^{qc}
	\approx
	\mathbb{T}_\sigma^q
	\sqrt{1+2
	c(\sigma_q)
	p}	
	,
\end{equation}
with
\begin{equation}\label{eq:T_approx_2}
	\mathbb{T}_\sigma^q
	=
	\frac{\mathbb{T}_q}{(1+\eta_\sigma\alpha_q)^2}
	\quad
	\text{and}
	\quad
	\mathbb{T}_q
	=
	\frac{4}{(Z_q)^2}
	\cdot
	\frac{\vFw}{v_0}
	,
\end{equation}
where $v_0=\sqrt{2\EF/\me}$. In the equations above,
$\mathbb{T}_\sigma^q$ represents the transmission coefficient of the $q$th interface [$q=L(\text{eft}),R(\text{ight})$] whose  spin-dependence stems exclusively from the spin selectiveness of the barrier. This effect will be analyzed in full detail in Sec.~\ref{sec:Spin_sel_bar}, and in the following discussion we assume spin non-selective barriers ($\alpha_q=0$). Interestingly, in such a case and for a small degree of spin-polarization of electrodes, one obtains 
$
	\mathcal{T}_\sigma^q
	\approx
	\mathbb{T}^q(1+\eta_\sigma p)
$.
Moreover, in the limit of weakly transparent barriers, 
$\mathcal{T}_\sigma^{qc}\ll1$,
and expanding $\cos\!\big(\theta_\sigma^c(\varepsilon)\big)$ around the resonant energy $\widetilde{\varepsilon}_{\sigma p}^c$, one finds that the expression for the transmission coefficient $\mathscr{T}_\sigma^c(\varepsilon)$ of the device takes the form of the Breit-Wigner formula~\cite{Stone_Phys.Rev.Lett.54/1985,Blanter_Phys.Rep.336/2000}
\begin{equation}\label{eq:Breit-Wigner}
	\mathscr{T}_\sigma^c(\varepsilon)
	=
	\mathscr{T}_{\text{max},\sigma}^c
	\frac{
		(\Gamma_\sigma^c)^2/4
	}{
	\big(\varepsilon-\widetilde{\varepsilon}_{\sigma p}^c\big)^2
	+
	(\Gamma_\sigma^c)^2/4
	}
	.
\end{equation}
In the equation above, 
$
	\Gamma_\sigma^c
	=
	\Gamma_\sigma^{Lc}+\Gamma_\sigma^{Rc}
$
and
$
	\mathscr{T}_{\text{max},\sigma}^c
	=
	4\Gamma_\sigma^{Lc}\Gamma_\sigma^{Rc}
	/
	(
	\Gamma_\sigma^c
	)^2
$
denotes the maximal value of the transmission coefficient at resonance, whereas 
$\Gamma_\sigma^{qc}=\hbar\nu\mathcal{T}_\sigma^{qc}$ is the decay width of the resonant level due to tunneling of electrons with spin $\sigma$ through the $q$th interface. It is expressed in terms of the attempt frequency $\nu$ defined as 
$
	\nu^{-1}
	=
	\hbar
	(\text{d}\theta_\sigma^c(\varepsilon)/\text{d}\varepsilon)|_{\varepsilon=\widetilde{\varepsilon}_{\sigma p}^c}
	=
	2\ell/\vFw
$~\cite{Price_Am.J.Phys.66/1998},
which basically describes the number of chances per unit time an electron that enters a CNT through the $q$th interface has to leave it through the same interface.

Using Eq.~(\ref{eq:Breit-Wigner}) together with Eq.~(\ref{eq:G_def}) one can then find the asymptotic values of TMR for large $Z$ to be:
(i) off resonance, i.e., when $\varepsilon-\widetilde{\varepsilon}_{\sigma p}^c\gg\Gamma_\sigma^c/2$, 
\begin{equation}\label{eq:TMR_off-res}
	\text{TMR}_\text{off-res}
	=
	\frac{
	1
	-
	\sqrt{1-4p^2}
	}{
	\sqrt{1-4p^2}
	}
	,
\end{equation}
which for $p=0.25$ yields $\text{TMR}_\text{off-res}\approx15.5$~\%;
(ii) at resonance, i.e., when $\varepsilon=\widetilde{\varepsilon}_{\sigma p}^c$, 
\begin{equation}
	\text{TMR}_\text{res}
	=
	\frac{1}{2}\,\text{TMR}_\text{off-res}
	.
\end{equation}
The variation of TMR between these two limiting values can be seen as a double-dotted-dashed line in Figs.~\ref{fig2}(c), where the dips correspond to resonant tunneling of electrons ---this also manifests as peaks in conductance given by double-dotted-dashed line in Fig.~\ref{fig2}(f). More numerical examples of $\text{TMR}_\text{off-res}$ for large $Z$ and different $p$ can bee seen in Fig.~\ref{fig3}(c). 
Furthermore, it is worth noting that if in derivation of the formula above instead of Eq.~(\ref{eq:T_approx}) one employs its counterpart for low spin polarizations of electrodes, the Julli\`{e}re value of tunneling magnetoresistance~\cite{Julliere_Phys.Lett.A54/1975}, $\text{TMR}_\text{off-res}=2p^2/(1-p^2)$, is recovered.

Another observation one can make is that the position of the resonances in conductance in Fig.~\ref{fig2}(d) is independent of $Z$ for large $Z$, whereas as $Z$ gets diminished their position becomes sensitive to $Z$. As already mentioned, this effect stems from the fact that when $Z$ increases the spin-dependent interfacial shifts $\varphi_\sigma^q$ for both spin orientations become equal at some point, and for even larger $Z$ they remain independent of the barrier strength, taking a constant value of $-\pi$, as can be seen in Fig.~\ref{fig3}(b). Furthermore, it is clear that for almost fully transparent (very small $Z$) and non-transparent (large $Z$) interfaces the part of the phase factor $\theta_\sigma^c(\varepsilon)$ in Eq.~(\ref{eq:TT_def}) corresponding to the spin-dependent interfacial phase shift is $\varphi_\sigma^{Lc}+\varphi_\sigma^{Rc}\approx-2\pi$, see Fig.~\ref{fig3}(b), regardless of the magnetic configuration of the spin valve. This is not the case for the intermediate regime of the barrier strength~$Z$, where $\varphi_\sigma^{Lc}+\varphi_\sigma^{Rc}<-2\pi$ and it is different for the parallel ($c=\text{P}$) and antiparallel ($c=\text{AP}$) magnetic configuration, so that the effect of spin-dependent backscattering of electrons into a CNT becomes visible in the TMR signal. For this reason, it is justified to neglect the spin-dependent interfacial phase shift for very small and large $Z$, and one can use this phase shift as an indication for an intermediate barrier strength ($0.1\lesssim Z \lesssim10$).

\begin{figure}[t]
	\includegraphics[scale=1]{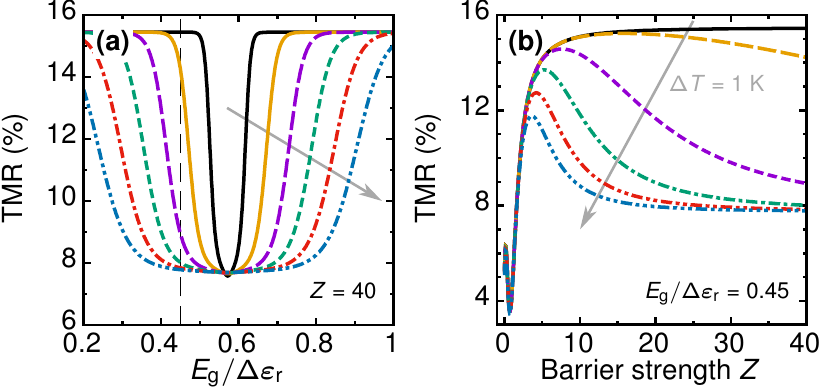}
	\caption{
	(color online)
	(a) Dependence of tunneling magnetoresistance (TMR) on temperature in the vicinity of the resonant transmission in the case of a large barrier strength ($Z=40$). The solid line corresponds to $T=1$~K and the gray arrow indicates the increase of temperature with the step of $\Delta T=1$~K up to 6~K for the double-dotted-dashed line.
	(b) Evolution of TMR as a function of the barrier strength $Z$ shown for different tempeatures and $\Eg/\Delta\varepsilon_\text{r}=0.45$, which is schamtically represented by a vertical dashed line in (a). 
	Note that the same pattern scheme for temperatures as in panel (a) is used.
	}
	\label{fig4}
\end{figure}

To complete the discussion of asymptotic values of TMR for large $Z$, we note that one should be careful when estimating the spin-polarization coefficient $p$ of electrodes. If one adjusts the gate voltage in such a way that the device is in the transport regime close to the resonant one but still off-resonant [compare dashed lines in Fig.~\ref{fig4}(a)], the TMR signal can become dependent on temperature, see Fig.~\ref{fig4}. In particular, the thermal broadening of the resonant peak in conductance leads also to a wider dip in TMR, as shown in Fig.~\ref{fig4}(a). When analyzing TMR as a function of the barrier strength $Z$, Fig.~\ref{fig4}(b), this, in turn, can be observed for large $Z$ as a thermally induced transition of TMR between the two limiting values discussed above. Since the period of the oscillations $\depsr$ is inversely proportional to the length $\ell$ of a CNT, one expects that such an effect of temperature on TMR to be more profound for longer CNTs.

\subsubsection{\label{sec:Asymm_bar}Asymmetry of tunnel barriers}

\begin{figure}[t]
	\includegraphics[scale=1]{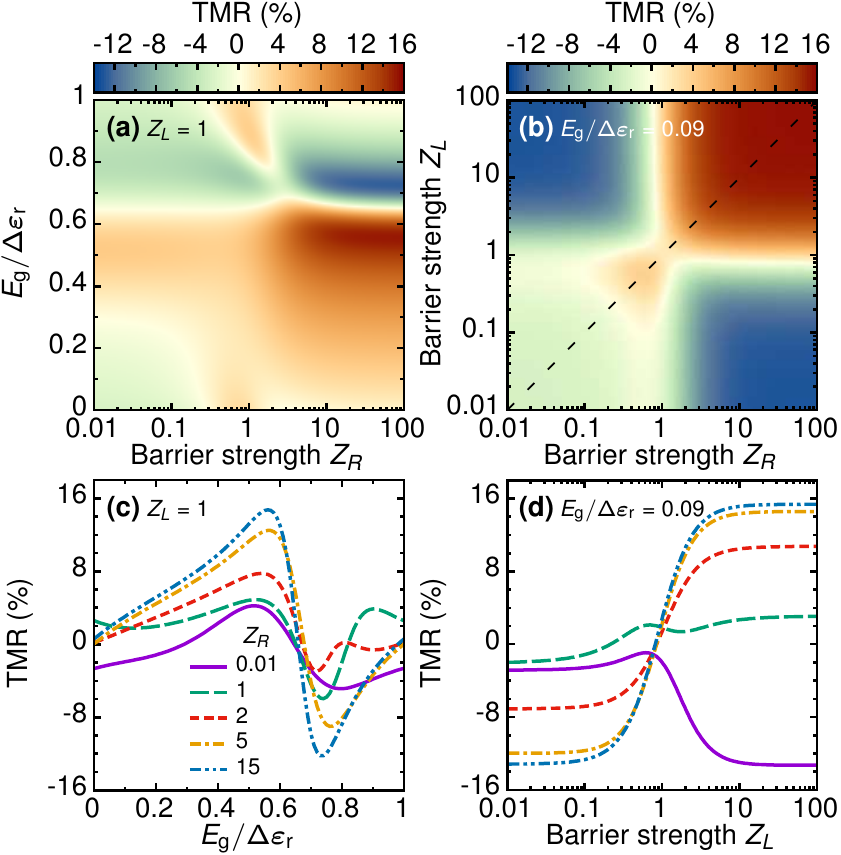}
	\caption{
	(color online)
	The effect of the left-right barrier strength asymmetry on tunneling magnetoresistance (TMR):
	(a) Analogous to Fig.~\ref{fig2}(a) except that at present the strength of only the right barrier $Z_R$ is changed over one period~$\depsr$ for $Z_L=1$.
	(b)~Density map of $\text{TMR}(Z_R,Z_L)$ for~$\Eg/\depsr\approx0.09$, which corresponds to the off-resonance limit for large~$Z$, and, in particular, to the position in the middle between two neighboring dips in TMR [see the double-dotted-dashed line in Fig.~\ref{fig2}(c)]; 
	Note that the bright color in (a)-(b) represents $\text{TMR}\approx0$, whereas the dashed line in (b) denotes the symmetric case of $Z_L=Z_R$. 
	\emph{Bottom panel} [(c)-(d)]:
	(c)~Cross-sections of (a) for indicated values of the right barrier strength~$Z_R$. 
	Here, the long-dashed (green) line represents the symmetric case ($Z_L=Z_R$), cf. Fig.~\ref{fig2}(c).
	(d)~Dependence of TMR on $Z_L$ shown for $\Eg/\depsr\approx0.09$ and values of $Z_R$ given in the legend of panel~(c) ---i.e., the vertical cross-sections of (b).
	All other parameters as in Fig.~\ref{fig2}.
	}
	\label{fig5}
\end{figure}

Let us now go beyond the assumption that both the tunnel barriers are identical, and consider the asymmetrical situation ($Z_L\neq Z_R$). 
This is illustrated in Fig.~\ref{fig5}(a), which in a similar fashion to Fig.~\ref{fig2}(a) presents the evolution of TMR in response to increasing now only the strength of the right barrier $Z_R$, while the strength of the left barrier is kept constant $Z_L=1$. Note that for the sake of clarity, only one period in $\Eg/\depsr$ has been plotted here. 
Noticeably, while for a vanishingly small right barrier ($Z_R\ll Z_L$) TMR remains qualitatively the same as in the case of the symmetric barriers, for the strong asymmetry of tunnel barriers, that is, $Z_R\gg Z_L$, a significant modification of TMR is observed. In particular, a distinctive saw-like pattern develops in this limit with large negative values of TMR, see Fig.~\ref{fig5}(c). In fact, such an asymmetry in the strength of tunnel barriers was essential to take into account in order to explain the occurrence of a negative TMR signal	in the experimental study of a spin-polarized transport through a CNT by Sahoo \emph{et al.}~\cite{Sahoo_NaturePhys.1/2005} ---see the lines for $Z_R=5$ (dotted-dashed) and $Z_R=15$ (double-dotted-dashed) in Fig.~\ref{fig5}(c) which qualitatively reproduces their result.

Next, to gain a better insight into how the asymmetry of the barriers affects TMR, in Fig.~\ref{fig5}(b) we show the dependence of TMR on the strength of both the left ($Z_L$) and right ($Z_R$) barriers for the gate-induced energy shift $\Eg/\depsr=0.09$ corresponding to the off-resonant limit from Fig.~\ref{fig2}(a).
The dashed line serves here merely as a guide for the eye denoting the case of identical barriers, with corresponding cross-sections along this line given by a solid  curve in Fig.~\ref{fig2}(b). Departing in either direction perpendicular to the dashed line represents the situation when one of the barriers increases whereas the other one gets smaller and smaller. A dramatic change in TMR occurs when one of the barriers becomes very small. Noticeably, TMR can take then large negative values which means that the device displays higher conductance in the antiparallel magnetic configuration of electrodes.

Employing the Breit-Wigner formula~(\ref{eq:Breit-Wigner}) for the situation when the strength of one tunnel barrier is significantly larger than the other one (i.e, asymmetric barriers, referred to as `as') and assuming, e.g., $Z_R\gg Z_L$ which corresponds to $\Gamma_\sigma^{Lc}\gg\Gamma_\sigma^{Rc}$ [recall that $\Gamma_\sigma^{qc}\propto\mathcal{T}_\sigma^{qc}$ and $\mathcal{T}_\sigma^{qc}\propto1/(Z_q)^2$, see Eqs.~(\ref{eq:T_approx})-(\ref{eq:T_approx_2})], we find the asymptotic form for the TMR at resonance,
\begin{equation}\label{eq:TMR_res_as}
	\text{TMR}_\text{res}^\text{as}
	=
	\sqrt{1-4p^2}-1
	,
\end{equation}
whereas the low-spin-polarization expression for the transmission coefficients of the barriers yields 
$
	\text{TMR}_\text{res}^\text{as}
	=
	-2p^2/(1+p^2)
$, 
in agreement with previous studies~\cite{Cottet_Sem.Sci.Tech.21/2006}.  On the other hand, in the off-resonant case the analogous asymptotic formula for $\text{TMR}_\text{off-res}^\text{as}$ is identical with Eq.~(\ref{eq:TMR_off-res}).
Importantly, we recall that these two asymptotic expressions for TMR are in general valid only if $Z_L,Z_R\gtrsim10$, that is, for weakly transparent barriers ($\mathcal{T}_\sigma^{qc}\ll1$), cf. Fig.~\ref{fig3}(a). Nevertheless, one can already see that the negative value of TMR in Fig.~\ref{fig5}(a) is very close to $\text{TMR}_\text{res}^\text{as}$, whereas in Fig.~\ref{fig5}(b) the asymptotic value $\text{TMR}_\text{off-res}^\text{as}$ is reached as soon as $Z_L,Z_R>2$ (see the top right corner of the plot).
As one can see in Fig.~\ref{fig5}(c), the tunability with respect to gate response of the TMR signal is strongest in the asymmetric case, if one barrier is very strong, here $Z_R=15$, while the strength of the second barrier assumes a value of about $Z_L=1$.

Concluding the results for barriers without spin selectivity, it is now clear that the largest TMR signal of $15.5$\% is obtained, if a device with realistic parameters, as specified at the beginning of Sec.~\ref{sec:Numerical_results}, is tuned to be off-resonant and if the tunnel barriers are strong ($Z\gtrsim 10$). Additionally, the response to a gate voltage is strongest, if the barriers are asymmetric with  (again tuned off-resonant). For instance, for $Z_L=1$ and $Z_R=15$ [see the double-dotted line in Fig.~\ref{fig5}(c)] and assuming a realistic gate coupling of $c_\text{gate}=0.33$, the TMR signal can be tuned from $\sim+14$\% to $\sim -12$\% within $5$~mV gate voltage.

Such devices can be fabricated using CoPd as ferromagnetic leads that mainly show low or intermediate tunnel barriers with $Z\lesssim 10$ (cf. Ref.~\cite{Morgan16}) adding a thin insulating layer between CNT and one contact (both contacts) for asymmetric (symmetric) barriers. If a spin selective insulator is used, the barriers will additionally become spin-selective.

\subsubsection{\label{sec:Spin_sel_bar}Spin-selective barriers}

Finally, we address the situation when tunnel barriers at the interfaces between electrodes and a CNT are additionally spin-selective, that is, $\alpha_L\neq0$ and/or $\alpha_R\neq0$. Such a situation can arise when spin selective insulators like EuO~\cite{Mueller09} or EuS~\cite{Nag07} or chiral molecules~\cite{Goehler11, Mishra13, Guo14} are used as tunnel barriers.
For the simplicity of the following discussion, we return to the situation of the symmetric barriers ($Z_L=Z_R=Z$), and only at the end of the section we consider the case of asymmetric barriers ($Z_L\neq Z_R$), which is expected to be more common for real devices. 

\begin{figure}[t]
	\includegraphics[scale=1]{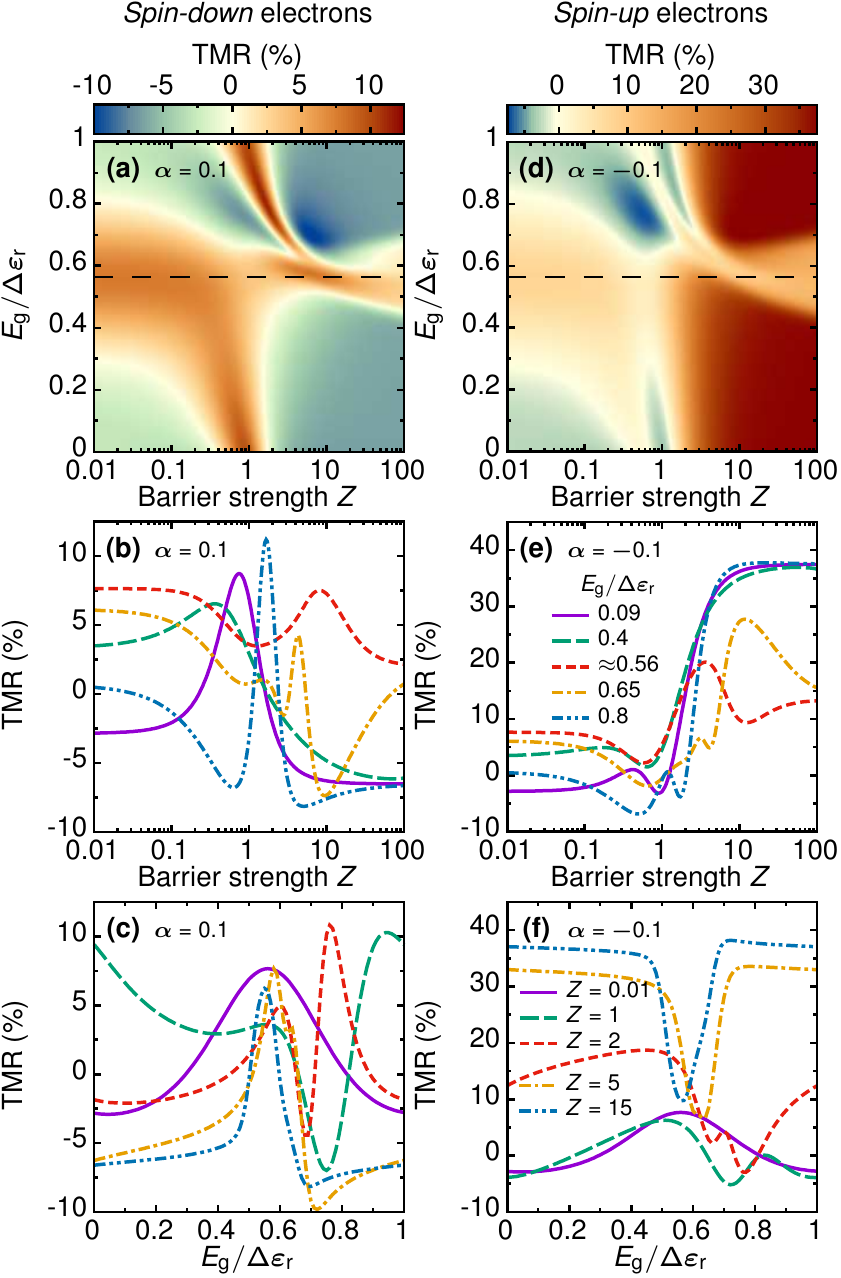}
	\caption{
		(color online)
		Analogous to the left panel [(a)-(c)] of Fig.~\ref{fig2} except that now it is assumed that tunnel barriers are spin selective ($\alpha_L=\alpha_R=\alpha$).
		Results for two different values of the barrier spin-asymmetry parameter $\alpha$ are shown: for $\alpha=0.1$ in left panel [(a)-(c)], when \emph{spin-down} electrons are preferred, and for $\alpha=-0.1$ in right panel [(d)-(f)], when \emph{spin-up} are preferred.
		Note that to facilitate the comparison of (a) and (d), the color scale in (a) is matching that of (d) in the corresponding range of TMR.
		Finally, we assume here again $Z_L=Z_R=Z$ and remaining parameters as in Fig.~\ref{fig2}.
	}
	\label{fig6}
\end{figure}

Numerical results illustrating how the spin-selectiveness of tunnel barriers affects the TMR are shown in Fig.~\ref{fig6} for identical barriers ($\alpha_L=\alpha_R=\alpha$).
Adding insulators between the CNT and the ferromagnetic leads will increase the barrier strength. Therefore, in our discussion we will focus on large tunnel barriers ($Z\gtrsim 10$). As visible in Figs.~\ref{fig6}(a) and~\ref{fig6}(d), cf. Fig.~\ref{fig2}(a),  tunnel barriers that filter incident electrons based on their spin orientation lead to significant, both qualitative and quantitative, changes in the TMR signal, which become especially visible for large barrier strength $Z$. Furthermore, this spin-filtering process, characterized by the spin asymmetry parameter~$\alpha$, Eq.~(\ref{eq:alpha}), depends essentially on whether more \emph{spin-up} [$\alpha<0$, as in Fig.~\ref{fig6}(d,e,f)] or \emph{spin-down} [$\alpha>0$, as in Fig.~\ref{fig6}(a,b,c)] electrons are passed through the barriers. Note that the spin orientation is defined with respect to the majority spins of the left electrode, which are defined as `spin-up' (cf. Fig.~\ref{fig1}).
Since the main quantitative difference between the two cases under discussion occurs in the limit of transitional and large~$Z$ [for small $Z$ there are neither qualitative nor quantitative differences between Figs.~\ref{fig6}(a) and~\ref{fig6}(d) ---mind the different scale ranges for the TMR], it may be instructive at this point to derive some asymptotic expressions for the TMR.

We use the Breit-Wigner formula, Eq.~(\ref{eq:Breit-Wigner}), to derive the following asymptotic expressions.
In the off-resonance limit for two symmetric  barriers (referred to by a superscript `s'), i.e., $\alpha_L=\alpha_R=\alpha$, one obtains
\begin{equation}\label{eq:TMR_off-res_s}
	\text{TMR}_\text{off-res}^\text{s}
	=
	\text{TMR}_\text{off-res}
	+
	\Delta\text{TMR}_\text{off-res}^\text{s}
\end{equation}
with
\begin{equation}\label{eq:corr_TMR_off-res_s}
	\Delta\text{TMR}_\text{off-res}^\text{s}
	=
	-
	\frac{
		4p
	}{
		\sqrt{1-4p^2}
	}
	\cdot
	\mathcal{F}_1(\alpha)
\end{equation}
and
\begin{equation}\label{eq:F1}
	\mathcal{F}_1(\alpha)
	=
	\frac{
	2\alpha\big(1+\alpha^2\big)
	}{
	4\alpha^2+\big(1+\alpha^2\big)^{\!2}
	}
	.
\end{equation}

On the other hand, at resonance for identical barriers one finds that

\begin{equation}
	\text{TMR}_\text{res}^\text{s}
	=
	\text{TMR}_\text{res}
	,
\end{equation}
which basically means that resonant transport of electrons through the device is insensitive to the spin-selectiveness of tunneling barriers, a fact that is discussed in more detail at the end of this section.

In general, if at least one barrier is spin-selective this leads to
%
%
a correction to the off-resonance $\text{TMR}$. This correction is determined  both by the spin-polarization of electrodes $p$ and by the spin-asymmetry of barriers~$\alpha$. 
What is more, the correction is positive~/~negative if spin-up ($\alpha<0$)~/~spin-down ($\alpha>0$) electrons are preferred. In the following, we assume the spin selectivity of the barriers to be $|\alpha|\leqslant0.25$, which is a very moderate choice regarding the fact that for EuO a spin filter efficiency as large as $80\%$ has been observed in tunnel junctions~\cite{Mueller09}. It can be checked that for $p=0.25$ one expects to achieve a TMR signal up to $\text{TMR}_\text{off-res}^\text{s}\approx37\%$ for strong barriers and spin-up electrons [see Fig.~\ref{fig6}(e)] and corrections as large as  $\Delta\text{TMR}_\text{off-res}^\text{s}\approx40\%$ compared to $\text{TMR}_\text{off-res}$.
Also, the gate-voltage response of the TMR signal is strongest for strong barriers [see Fig.~\ref{fig6}(f)], and tuning between $10\%$ and $37\%$ within a gate voltage of $5$ mV, assuming again gate coupling $c_\text{gate} = 0.33$. In contrast to spin-up electrons, the maximum value as well as the strongest gate response for the TMR signal for a spin-selective barrier that prefers spin-down electrons are in total not only smaller, but also found for small or intermediate barrier strength [see Fig.~\ref{fig6}(b) and~(c)]. Importantly, note that the spin moment of EuS aligns antiferromagnetically with respect to the spin moment of Co in Co/EuS multilayers~\cite{Pappas13}. For this reason, using EuS as spin-selective barrier with ferromagnetic leads from CoPd will most likely lead to a selection of spin-down electrons.

Though the fabrication of such a device is more tedious compared to symmetric barriers, it is possible to have only one spin-selective barrier $q$, i.e., $\alpha_L=\alpha$ and $\alpha_R=0$ for $q=L$ or $\alpha_L=0$ and $\alpha_R=\alpha$ for $q=R$, and in the off-resonant case one obtains
\begin{equation}\label{eq:TMR_off-res_q}
	\text{TMR}_\text{off-res}^{q}
	=
	\text{TMR}_\text{off-res}
	+
	\Delta\text{TMR}_\text{off-res}^q
\end{equation}
with
\begin{equation}\label{eq:corr_TMR_off-res_q}
	\Delta\text{TMR}_\text{off-res}^q
	=
	-
	\frac{
		4p
	}{
	\sqrt{1-4p^2}
	}
	\cdot
	\mathcal{F}_2(\alpha)
\end{equation}
and
\begin{equation}\label{eq:F2}
	\mathcal{F}_2(\alpha)
	=
	\frac{\alpha}{1+\alpha^2}
	.
\end{equation}

Clearly, only the dependence on $\alpha$ is affected by whether one or both barriers are spin-selective, cf. Eqs~(\ref{eq:corr_TMR_off-res_s}) and~(\ref{eq:corr_TMR_off-res_q}). For $\alpha\neq0$ and $|\alpha|<1$ one gets $|\mathcal{F}_1(\alpha)|>|\mathcal{F}_2(\alpha)|$, and the change in the TMR signal is reduced to $\Delta\text{TMR}_\text{off-res}^{L/R}\approx25\%$.

However, if only one barrier is spin selective, also the resonant TMR signal is changed:
\begin{equation}
	\text{TMR}_\text{res}^{q}
	=
	\text{TMR}_\text{res}
	+
	\Delta\text{TMR}_\text{res}^q
	,
\end{equation}
where
\begin{equation}
	\Delta\text{TMR}_\text{res}^q
	=
	\frac{1}{2}
	\big[\mathcal{F}_3(\alpha)-1\big]
	+
	\frac{
		\mathcal{F}_3(\alpha)
		\mathcal{F}_4^q(\alpha,p)-1
	}{
		2\sqrt{1-4p^2}
	}
\end{equation}
with
\begin{equation}
	\mathcal{F}_3(\alpha)
	=
	\frac{
		16+4\alpha^2\big[\alpha^2(1+\alpha^2)-4\big]
	}{
	\big(4+\alpha^4\big)^{\!2}
	}
\end{equation}
and
\begin{widetext}
\begin{equation}
	\mathcal{F}_4^{L/R}(\alpha,p)
	=
	\frac{
		2\sqrt{1-4p^2}
		\big(1+\alpha^2\mp 4p\alpha\big)
		\!
		\Big[1+\big(1-\alpha^2\big)^{\!2}\Big]
		+
		\big(1-4p^2\big)
		\!
		\Big[1+\big(1-\alpha^2\big)^{\!4}\Big]
		+
		\!
		\sum\limits_{\eta=\pm}
		\!\!
		\big(1+\eta 2p\big)^{\!2}
		\big(1\mp\eta \alpha\big)^{\!4}
	}{
	2\big(1+\alpha^2\mp 4p\alpha\big)
	\!
	\Big[1+\big(1-\alpha^2\big)^{\!2}\Big]
	+
	4\sqrt{1-4p^2}
	\big(1-\alpha^2\big)^{\!2}
}
.
\end{equation}
\end{widetext}
Note that $\lim\limits_{\alpha\rightarrow 0}\mathcal{F}_3(\alpha)=1$ and 
$\lim\limits_{\alpha\rightarrow 0}\mathcal{F}_4^{L/R}(\alpha,P)=1$, so that in the limit of vanishingly small spin-selectiveness of barriers we recover the previously found result, that is, $\text{TMR}_\text{res}^{q}=\text{TMR}_\text{res}$.

\begin{figure}[t]
	\includegraphics[scale=1]{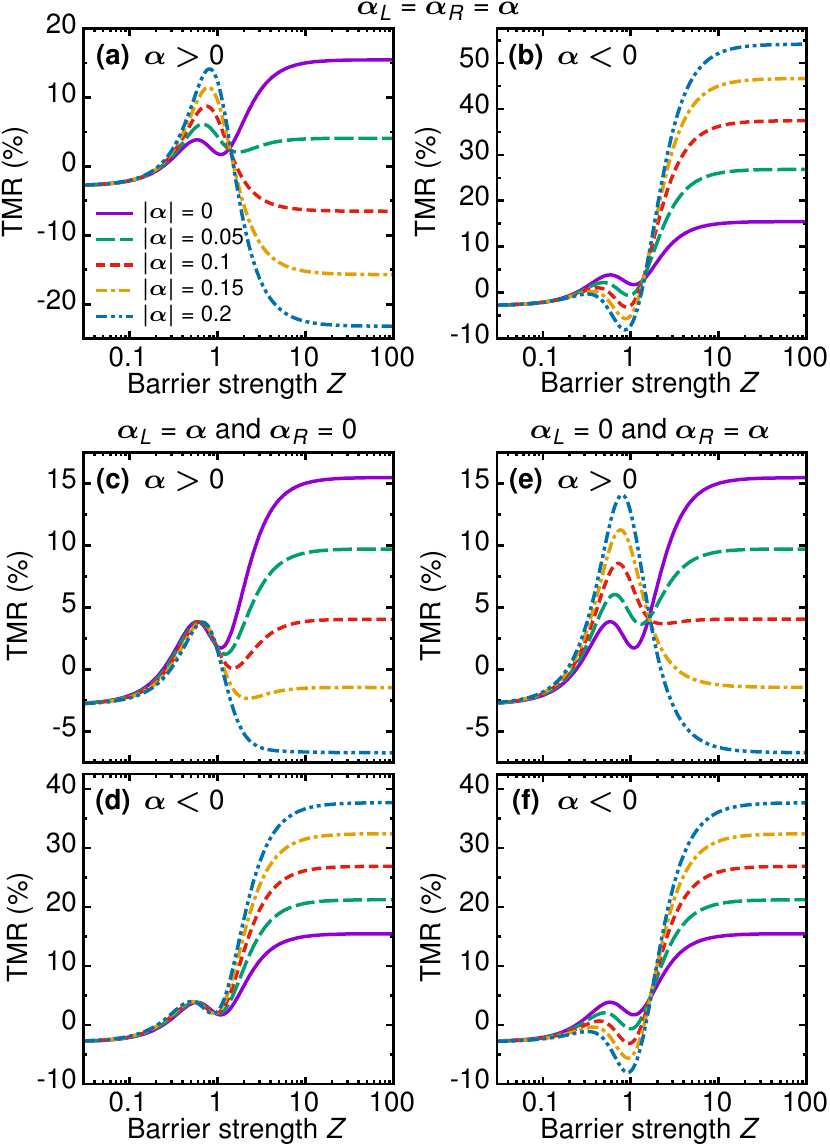}
	\caption{
		(color online)
		The effect of spin-selectivity of tunnel barriers on tunneling magnetoresistance (TMR) shown as a function of the barrier strength $Z$ for $\Eg/\depsr=0.09$ and several values of the spin asymmetry parameter $\alpha$. 
		\emph{Top panel} [(a)-(b)] represents the situation of identical barriers ($\alpha_L=\alpha_R=\alpha$), whereas in the \emph{bottom panel} [(c)-(f)] the case of only one barrier being spin-selective is shown, that is, the left one ($\alpha_L=\alpha$ and $\alpha_R=0$) in (c)-(d) and the right one ($\alpha_L=0$ and $\alpha_R=\alpha$) in (e)-(f). 
		Note that the solid line corresponds to $\alpha=0$ and serves as the reference line for comparison of different plots. 
		{This line is also identical to the solid line in Fig.~\ref{fig2}(b).}
		All other parameters as in Fig.~\ref{fig2}.
	}
	\label{fig7}
\end{figure}

Figure~\ref{fig7} presents the evolution of the off-resonance TMR as a function of the barrier strength $Z$ for selected values of the spin asymmetry parameter $\alpha$ in three specific cases: (a)-(b) when both tunnel barriers are identical ($\alpha_L=\alpha_R$), or when only one of the barriers is spin-selective: \emph{left} in (c)-(d) and \emph{right} in (e)-(f). There is no dependence of TMR on $\alpha$ seen for small $Z$, whereas for large $Z$ a significant variation of TMR occurs, with the asymptotic values of TMR given by the expressions above. Moreover, in the latter limit one observes a general trend that for positive $\alpha$ (spin-down electrons preferred) TMR becomes decreased, so that for sufficiently large $\alpha$ it can get negative, whereas for negative $\alpha$ (spin-up electrons preferred) TMR increases.
Interestingly, for the transitional values of $Z$ we find that TMR varies non-monotonically in the case of identical barriers and only the right barrier being spin-selective. On the other hand, in  the case of only the left barrier spin-selective TMR remains rather unaffected by $\alpha\neq0$ and only as $Z$ is further increased TMR starts gradually approaching its asymptotic values. As previously, this behavior can be understood in terms of spin-dependent transmission coefficient and interfacial phase shift for a single tunnel barrier. Importantly, if only the left barrier is spin-selective, its effect is the same for both magnetic configurations of electrodes, so that the TMR is only slightly influenced. This is due to the fact that the orientation of the spin moment of left electrode defines here the reference frame. The situation is different when the right barrier is spin-selective. In such a case, depending on the magnetic configuration the barrier prefers either spin-up or spin-down electrons and thus, conductances in both magnetic configurations are affected differently, which ultimately reveals itself in the TMR signal. 

Finally, we note that in real devices one should in general expect that the combination of the two effects studied above will occur, that is, the two tunnel barriers will be asymmetric both in terms of strength ($Z_L\neq Z_R$)  and spin selectiveness ($\alpha_L\neq\alpha_R$). We find that in such a case the previously derived asymptotic formulae for strongly asymmetric barriers ($Z_R\gg Z_L$), see Sec.~\ref{sec:Asymm_bar}, become modified as follows to incorporate the effect of different spin-selective properties of each barrier (we use a prime to distinguish this case):
off resonance one obtains 
\begin{equation}\label{eq:TMR_off-res_as_ss}
	\big(\text{TMR}_\text{off-res}^\text{as}\big)^\prime
	=
	\text{TMR}_\text{off-res}
	+
	\Delta\text{TMR}_\text{off-res}^\text{as}
\end{equation}
with
\begin{equation}
	\Delta\text{TMR}_\text{off-res}^\text{as}
	=
	-
	\frac{
	4p
	}{
	\sqrt{1-4p^2}
	}
	\cdot
	\mathcal{S}_+(\alpha_L,\alpha_R)
	,
\end{equation}
whereas at resonance one gets
\begin{equation}\label{eq:TMR_res_as_ss}
	\big(\text{TMR}_\text{res}^\text{as}\big)^\prime
	=
	\frac{
	\text{TMR}_\text{res}^\text{as}
	}{
	1-4p\mathcal{S}_-(\alpha_L,\alpha_R)
	}
	+
	\Delta\text{TMR}_\text{res}^\text{as}
\end{equation}
with
\begin{equation}
	\Delta\text{TMR}_\text{res}^\text{as}
	=
	\frac{
	4p\mathcal{S}_-(\alpha_L,\alpha_R)
	}{
	1-4p\mathcal{S}_-(\alpha_L,\alpha_R)
	}
	.
\end{equation}
The coefficient $\mathcal{S}_\pm(\alpha_L,\alpha_R)$, defined as 
\begin{equation}
	\mathcal{S}_\pm(\alpha_L,\alpha_R)
	=
	\frac{
	(\alpha_L\pm\alpha_R)(1\pm\alpha_L\alpha_R)
	}{
	(\alpha_L\pm\alpha_R)^2+(1\pm\alpha_L\alpha_R)^2
	}
	,
\end{equation}
describes the asymmetry of tunnel barriers due to difference in spin asymmetry parameters between left ($\alpha_L$) and right ($\alpha_R$) barrier. 
One can then notice that for the symmetric case, that is, when $\alpha_L=\alpha_R=\alpha$, one obtains $\mathcal{S}_+(\alpha,\alpha)\equiv\mathcal{F}_1(\alpha)$, see Eq.~(\ref{eq:F1}), so that asymptotic  equations for TMR given by Eqs.~(\ref{eq:TMR_off-res_s}) and~(\ref{eq:TMR_off-res_as_ss}) become identical.
Similarly, one finds the relation between Eqs.~(\ref{eq:TMR_off-res_q}) and~(\ref{eq:TMR_off-res_as_ss}) for only a single barrier being spin-selective, $\mathcal{S}_+(\alpha,0)=\mathcal{S}_+(0,\alpha)\equiv\mathcal{F}_2(\alpha)$, see Eq.~(\ref{eq:F2}). 
The analysis of $\mathcal{S}_+(\alpha_L,\alpha_R)$ brings us to a conclusion that $\left(\text{TMR}_\text{off-res}^\text{as}\right)^\prime$ can be effectively maximized by ensuring that the barriers are symmetric ($\alpha_L=\alpha_R$) and engineering them in such a way that \emph{spin-up} electrons are favored (i.e., $\alpha_L,\alpha_R<0$).

On the other hand, in the resonant case we notice that if both barriers are identical ($\alpha_L=\alpha_R=\alpha$), the spin-selectiveness of barriers plays no role, as $\mathcal{S}_-(\alpha,\alpha)=0$ and Eq.~(\ref{eq:TMR_res_as}) is recovered.
This striking difference can be qualitatively understood by considering how the spin-selectiveness of barriers affects conductance. In the case of strongly asymmetric barriers under discussion, one finds that the spin-resolved conductance in the magnetic configuration  $c=\text{P,AP}$ depends on transmission coefficients~(\ref{eq:T_approx}) of left ($\mathcal{T}_\sigma^{Lc}$) and right ($\mathcal{T}_\sigma^{Rc}$) barriers approximately as
\begin{equation}
	\big[G_\sigma^c\big]_\text{off-res}
	\propto
	\mathcal{T}_\sigma^{Lc}\mathcal{T}_\sigma^{Rc}
	\quad
	\text{and}
	\quad
	\big[G_\sigma^c\big]_\text{res}
	\propto
	\frac{\mathcal{T}_\sigma^{Rc}}{\mathcal{T}_\sigma^{Lc}}
	.
\end{equation}
Consequently, one can see that for resonant transport contributions due to the spin-selectivity of barriers cancel each other if these exhibit identical properties in terms of spin-dependent transparency.
Interestingly, by optimizing the barriers one also expects to observe positive $\big(\text{TMR}_\text{res}^\text{as}\big)^\prime$ in the resonant transport case, which is generically negative as given by Eq.~(\ref{eq:TMR_res_as}). This can be achieved by forcing $\alpha_L>\alpha_R$ with a further constraint put on $\alpha_L$ determined by the value of $p$. Large positive values of $\big(\text{TMR}_\text{res}^\text{as}\big)^\prime$ are especially expected for $\alpha_L>0$, which means that the left barrier should favor minority (\emph{spin-down}) electrons.  For instance, let us assume that only the left barrier is modified to be spin-selective, that is, $\alpha_L\equiv\alpha$ and $\alpha_R=0$. We find numerically (for $p=0.25$) that $\big(\text{TMR}_\text{res}^\text{as}\big)^\prime>0$ as soon as $\alpha>\alpha_0$ with $\alpha_0\approx0.14$, and the  increase of $\alpha$ is followed by the monotonic growth of $\big(\text{TMR}_\text{res}^\text{as}\big)^\prime$ up to a value of $\approx73\%$ for $\alpha=1$ ---the maximal achievable value for given~$p$. Interestingly, if one could fabricate a device with $\alpha_L=-\alpha_R=\alpha$, that is, with the tunnel barriers of perfectly antisymmetric spin-selective properties, this would allow for achieving $\alpha_0\approx0.07$ and $\big(\text{TMR}_\text{res}^\text{as}\big)^\prime\gtrsim50\%$ already at $\alpha=0.3$.
%

\subsection{\label{sec:Many_channels}The case of many orbital channels}

In this section we relax the assumption regarding the position of the Fermi level around the charge neutrality point (i.e., $\EFw=0$), and assume that the level has been shifted, see the right side of Fig.~\ref{fig1}(b). For illustrative purposes, we consider two cases of $\EFw=400$~meV and $\EFw=650$~meV, which means that 2 ($n=0,1$) and 3 ($n=0,1,2$) orbital channels (subbands), respectively, are available for charge and spin transport through the device.

\begin{figure}[t]
	\includegraphics[scale=1]{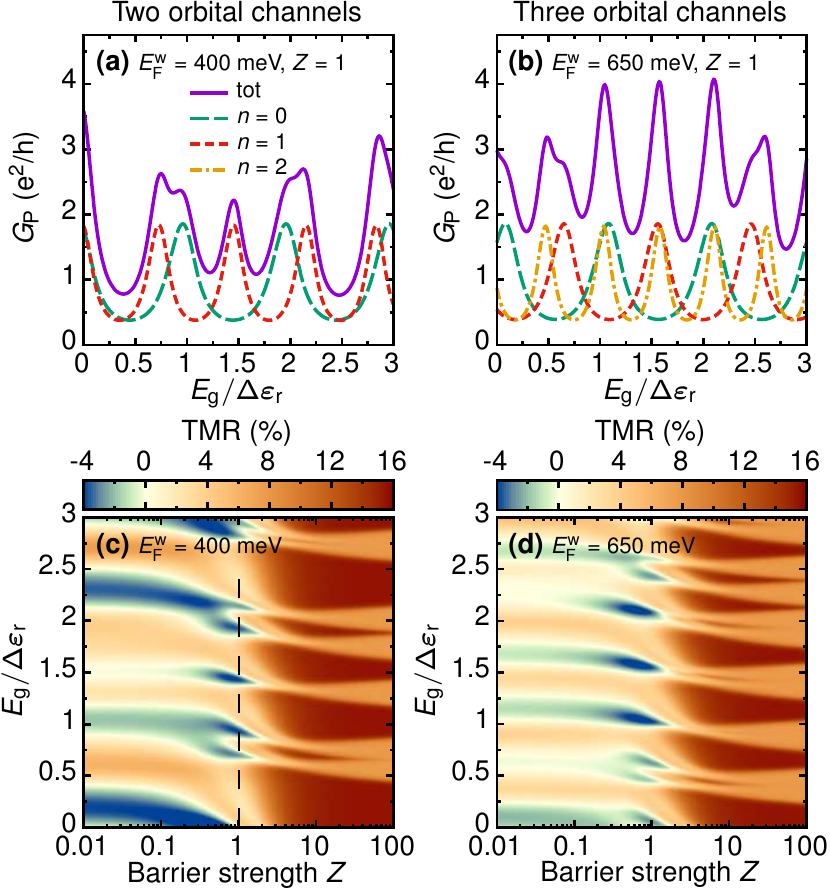}
	\caption{
		(color online)
		The effect of many orbital channels on transport properties of a CNT-based spin valve. The number of such channels participating in transport is modified here by assuming different shifts of the Fermi level $\EFw$: in the \emph{left column} $\EFw=400$~meV (two orbital channels included), whereas in the \emph{right column} $\EFw=650$~meV (three orbital channels included).
		\emph{Top panel} [(a)-(b)]: Conductance $G_\text{P}$ in the parallel magnetic configuration (solid line) decomposed into contributions from different orbital channels (dashed and dotted-dashed lines) shown as a function of the shift of the Fermi level $\Eg$ and $Z_L=Z_R=Z=1$.
		\emph{Bottom panel} [(c)-(d)]: Density map of tunneling magnetoresistance (TMR) plotted as a function of $\Eg$ and the barrier strength $Z$. Note that cross-sections of (c) for $Z=1$ (that is, along the horizontal thin dashed line) and $Z=100$ are shown in Fig.~\ref{fig9}(b).
		All other parameters as in Fig.~\ref{fig2}.
	}
	\label{fig8}
\end{figure}

The key difference with respect to the single-channel case stems from the fact that now conductance~$G_\text{P/AP}$, Eq.~(\ref{eq:G_def}), for each magnetic configuration has to be summed over all orbital transport channels. Since each  channel is described by a different transmission coefficient~$\mathscr{T}_{\sigma n}^\text{P/AP}$, Eq.~(\ref{eq:TT_def}), characteristic energies $\widetilde{\varepsilon}_{\sigma p}^{cn}$ at which resonant tunneling of electrons occurs are uniquely associated with the subband index $n$, 
\begin{multline}\label{eq:res_position_n}
	\widetilde{\varepsilon}_{\sigma p}^{cn}
	=
	\depsr
	\sqrt{
	\Big(
	p
	-
	\dfrac{1}{2\pi}
	\big(
	\varphi_{\sigma n}^{Lc}
	+
	\varphi_{\sigma n}^{Rc}
	\big)
	-
	\frac{\ell\kFw}{\pi}
	\Big)^2
	+
	\Big(
	\frac{\ell n}{r\pi}
	\Big)^2
	}
	\\
	+
	\Eg
	-
	\EFw
	,
\end{multline}
for $p\in\mathbb{Z}$. Consequently, resonances in conductance for channels characterized by various $n$ appear at different intervals, which, in turn, leads to a complex pattern of total conductance as a function of $\Eg$. This effect is illustrated in the top panel of Fig.~\ref{fig8}, where, as an example, the total conductance in the parallel magnetic configuration (solid line) for two (a) and three (b) orbital channels participating in transport has been decomposed into contributions from specific channels. 
Furthermore, the resultant TMR no longer exhibits a clear periodic pattern, see the bottom panel of Fig.~\ref{fig8}, where the $\Eg$-range is purposely assumed the same as in Fig.~\ref{fig2}(a) to enable easy comparison of the results. Nevertheless, one can still distinguish three distinctive regions with respect to the barrier strength $Z$, whose origin can be explained analogously as in the single-channel case, see Sec.~\ref{sec:Single_channel}. Importantly, it should be noticed that in the limit of large $Z$ the TMR varies between two characteristic values $\text{TMR}_\text{off-res}$, Eq.~(\ref{eq:TMR_off-res}), and $\text{TMR}_\text{res}=\text{TMR}_\text{off-res}/2$, corresponding to the off-resonant and resonant electron tunneling through a CNT, respectively. As the number of orbital channels participating in transport increases, also the chance of resonant tunneling becomes larger, because each channel has its own unique set of resonant energies~(\ref{eq:res_position_n}). As a result, one expects that with increasing channel number the TMR should take a resonant value more often, as observed comparing Figs.~\ref{fig8}(c) and~\ref{fig8}(d).

\begin{figure}[t]
	\includegraphics[scale=1]{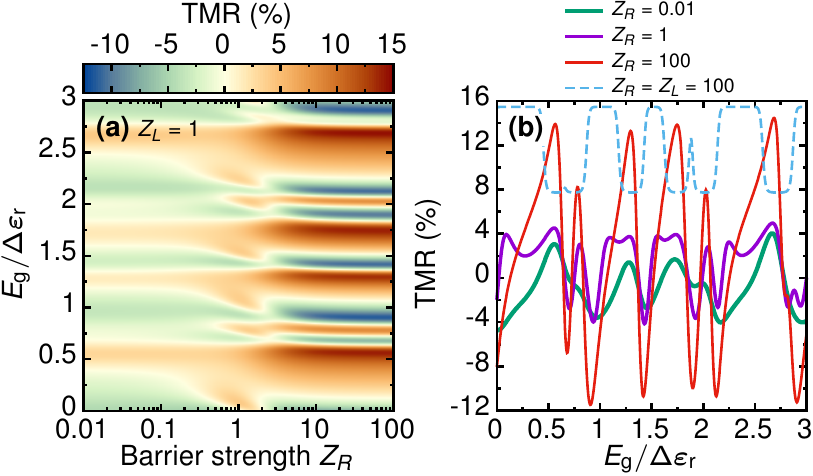}
	\caption{
		(color online)
		Influence of the barrier strength asymmetry on tunneling magnetoresistance (TMR). Assuming a fixed value of the left barrier strength $Z_L=1$, the change of TMR as a function of the right barrier strength $Z_R$ is presented in (a).
		Characteristic cross-sections of (a) for selected  values of $Z_R$ are shown in (b). Note that the case of $Z_R=1$ corresponds to symmetric barriers ($Z_L=Z_R$), and, thus, it also represents the cross-section of Fig.~\ref{fig8}(c) along the horizontal thin dashed line.
		Moreover, the result for asymptotically large, symmetric barriers ($Z_L=Z_R=100$) is also shown (the dashed line).
		Here, we assume $\EFw=400$~meV and two orbital channels are taken into account, whereas remaining parameters are as in Fig.~\ref{fig2}.
	}
	\label{fig9}
\end{figure}

Next, we analyze how the asymmetry of the strength between left and right barriers ($Z_L\neq Z_R$) affects the TMR signal. For this purpose, we assume that the left barrier is fixed with $Z_L=1$ and we alter the strength of right barrier $Z_R$, see Fig.~\ref{fig9}. The cross-section of (a) along $Z_R=1$ corresponds then to the cross-section along a thin dashed line in Fig.~\ref{fig8}(c), and represents the case of symmetric barriers. 
For $Z_R\ll1$, which represents the situation of the right barrier being almost fully transparent, one can see softening of TMR features which is accompanied by a smearing out of some resonances, see the relevant lines in Fig.~\ref{fig9}(b).
On the other hand, in the opposite limit ($Z_R\gg1$), that is, for a strong asymmetry between the barriers with the right barrier of vanishingly small transmission, TMR features become generally much sharper, forming a saw-like pattern, and TMR values vary in a broader range. Interestingly, it can be noticed that peaks and dips developing in TMR evolve from the same features which survive also in the low $Z_R$ limit. In addition, an especially stark contrast between symmetric (dashed line) and asymmetric (thin solid line) tunnel barriers is seen in the $Z_R$ limit under consideration.

\begin{figure}[t]
	\includegraphics[scale=1]{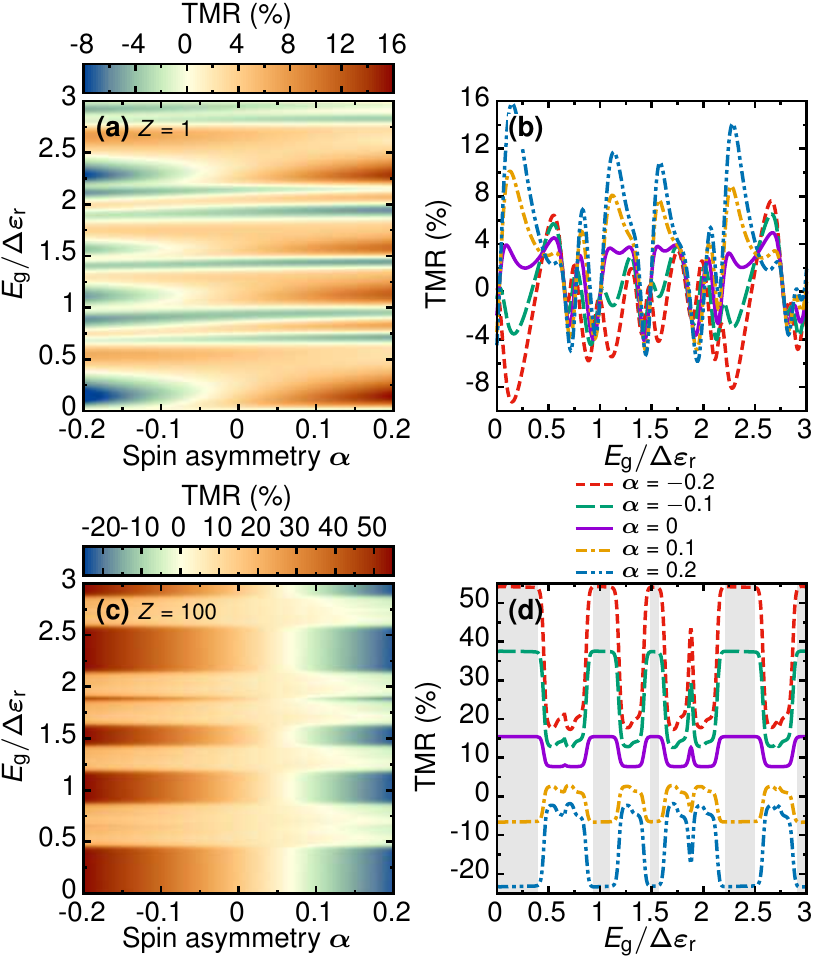}
	\caption{
		(color online)
		The effect of spin-selective barriers on tunneling magnetoresistance (TMR) in the case of two identical barriers characterized by the same strength ($Z_L=Z_R=Z$) and the spin asymmetry parameter $\alpha_L=\alpha_R=\alpha$. 
		\emph{Top panel} [(a)-(b)] represents the results for $Z=1$, whereas the \emph{bottom} one [(c)-(d)] for $Z=100$. Plots in the right column contain selected cross-sections of respective density plots from the left column. 
		Solid lines in (b) and (d) correspond to the situation of barriers being spin-non-selective ($\alpha=0$). Note, additionally, that the solid line in (b) [(d)] is identical with the solid line for $Z_R=1$ [dashed line] in Fig.~\ref{fig9}(b).
		All other parameters as in Fig.~\ref{fig9}.
	}
	\label{fig10}
\end{figure}

Finally, to make the present discussion complete, we also investigate the effect of spin-selective barriers. Since this aspect has been extensively analyzed in Sec.~\ref{sec:Spin_sel_bar} for the case of a single orbital channel, here we focus only on a specific situation of two identical barriers, that is, when $Z_L=Z_R=Z$ and $\alpha_L=\alpha_R=\alpha$. In Fig.~\ref{fig10} we show the evolution of TMR as a function of the spin asymmetry parameter $\alpha$ and the shift of the Fermi level $\Eg$ for two representative values of the barrier strength: $Z=1$ (a) and $Z=100$ (c), with selected cross-sections for chosen values of $\alpha$ given in (b) and (d), respectively. It can be seen that additional spin filtering of electrons by tunnel barriers can substantially modify the observed TMR. In the limit of large $Z$, illustrated in the bottom panel of Fig.~\ref{fig10}, it can be noticed that for the off-resonance regions, marked in (d) as shaded areas, the TMR suffers significant changes when $\alpha$ is appreciably large, while in the resonant regions the observed variation of the TMR effect is more moderate. Moreover, in the former case the dependence of TMR on $\alpha$ is described by Eq.~(\ref{eq:TMR_off-res_s}), exactly then same as in the situation of a single orbital channel.

\section{\label{sec:Conclusions}Conclusions}

With this paper we provide a complete physical picture of the TMR effect in CNT-based spin valves. In particular, we focus on the influence of the tunnel barrier strength and spin-selectivity of the barrier on the TMR. 
The largest TMR signals are generally found in the strong barrier case when the device is tuned to be off-resonant with regard to the Fabry-P\'erot resonances in the one-dimensional wire. For a realistic CNT based spin valve we find a TMR signal of $15.5\%$, a value we realized in an recent experiment~\cite{Morgan16}. In general, the off-resonant TMR is more sensitive toward changes in the barriers that the on-resonant TMR. For instance, the off-resonant TMR increases by $\Delta \text{TMR}_\text{off-res} = 40\%$ if spin-selective barriers are added that prefer majority (spin-up) electrons from electrode, while the resonant TMR signal does not change at all. Such a spin-selective barrier might be implemented by spin-selective insulators as EuS or EuO. However, these materials are likely to couple antiferromagnetically to the ferromagnetic leads. Therefore, using spin selecting molecules as barrier is believed to be more promising with regard to enhancing the TMR signal, especially since a moderate selectivity of $\sim10\%$ already yields a strong enhancement of the TMR signal up to $37\%$ and double stranded DNA has been shown to exhibit high spin filter efficiency~\cite{Xie11}. Using DNA as spin filter will require perpendicular orientation of the magnetization of the contacts. This can be implemented by the right choice contact material and contact shape. 

As shown before, the barrier strength in CNT spin valves can be asymmetric due to fabrication resulting in negative values of the TMR signal~\cite{Sahoo_NaturePhys.1/2005}. We find that it is in principle possible to correct this, if the barrier of the injection contact favors minority  (spin-down) electrons leading to large positive TMR of up to 50-70$\%$. In this case, adding an insulating of EuS or EuO to the lead used for spin injection will likely yield the desired result.

In the case of intermediate barrier strength, i.e., the potential energy of the barrier matches the energy of the incident electrons at the Fermi level, we show that the magnitude of the TMR has a strong response to the gate voltage varying from $+14\%$ to $-12\%$ within 5 mV gate voltage for a realistic device and without spin-selective barriers. It is important to note that this tunability of the TMR signal is effective in the absence of spin-orbit coupling, thus preserving the long spin relaxation time inherent for carbon materials. The tunability of the TMR signal is strongest for asymmetric barriers. Adding more transport channels, e.g., by working at larger gate voltages, the number of resonances increases leading to a less periodic pattern of the TMR with gate voltage. Changes of the TMR signal with respect to barrier strength, asymmetry and spin-selectivity, however, remain qualitatively the same.

In conclusion, we showed that the feasibility of modification of the tunnel barriers in a controlled way together with electrical tuning of a CNT  could open up a possibility to built CNT-based devices exhibiting large TMR effect with strong response to the gate voltage. Specifically, a prospective way to achieve this goal lies in application of highly asymmetric and/or spin-selective tunnel barriers. This paves the way for spintronic devices that work without spin-orbit coupling and thus preserve long spin relaxation times.

\acknowledgments

The authors thank P. Mavropoulos and J. Splettstoesser for fruitful discussions.
M.M. acknowledges financial support from the Alexander von Humboldt Foundation, the Polish Ministry of Science and Higher Education through a young scientist fellowship (0066/E-336/9/2014), and the Knut and Alice Wallenberg Foundation. C. Meyer acknowledges financial support from the DFG Research unit FOR912 and by the ``Nieders\"achsiche Vorab'' program of the Volkswagen Stiftung.



\begin{thebibliography}{68}%
\makeatletter
\providecommand \@ifxundefined [1]{%
 \@ifx{#1\undefined}
}%
\providecommand \@ifnum [1]{%
 \ifnum #1\expandafter \@firstoftwo
 \else \expandafter \@secondoftwo
 \fi
}%
\providecommand \@ifx [1]{%
 \ifx #1\expandafter \@firstoftwo
 \else \expandafter \@secondoftwo
 \fi
}%
\providecommand \natexlab [1]{#1}%
\providecommand \enquote  [1]{``#1''}%
\providecommand \bibnamefont  [1]{#1}%
\providecommand \bibfnamefont [1]{#1}%
\providecommand \citenamefont [1]{#1}%
\providecommand \href@noop [0]{\@secondoftwo}%
\providecommand \href [0]{\begingroup \@sanitize@url \@href}%
\providecommand \@href[1]{\@@startlink{#1}\@@href}%
\providecommand \@@href[1]{\endgroup#1\@@endlink}%
\providecommand \@sanitize@url [0]{\catcode `\\12\catcode `\$12\catcode
  `\&12\catcode `\#12\catcode `\^12\catcode `\_12\catcode `\%12\relax}%
\providecommand \@@startlink[1]{}%
\providecommand \@@endlink[0]{}%
\providecommand \url  [0]{\begingroup\@sanitize@url \@url }%
\providecommand \@url [1]{\endgroup\@href {#1}{\urlprefix }}%
\providecommand \urlprefix  [0]{URL }%
\providecommand \Eprint [0]{\href }%
\providecommand \doibase [0]{http://dx.doi.org/}%
\providecommand \selectlanguage [0]{\@gobble}%
\providecommand \bibinfo  [0]{\@secondoftwo}%
\providecommand \bibfield  [0]{\@secondoftwo}%
\providecommand \translation [1]{[#1]}%
\providecommand \BibitemOpen [0]{}%
\providecommand \bibitemStop [0]{}%
\providecommand \bibitemNoStop [0]{.\EOS\space}%
\providecommand \EOS [0]{\spacefactor3000\relax}%
\providecommand \BibitemShut  [1]{\csname bibitem#1\endcsname}%
\let\auto@bib@innerbib\@empty
\bibitem [{\citenamefont {Bader}\ and\ \citenamefont
  {Parkin}(2010)}]{Bader_Annu.Rev.Condens.MatterPhys.1/2010}%
  \BibitemOpen
  \bibfield  {author} {\bibinfo {author} {\bibfnamefont {S.~D}\ \bibnamefont
  {Bader}}\ and\ \bibinfo {author} {\bibfnamefont {S.~S.~P.}\ \bibnamefont
  {Parkin}},\ }\bibfield  {title} {\enquote {\bibinfo {title} {Spintronics},}\
  }\href@noop {} {\bibfield  {journal} {\bibinfo  {journal} {Annu. Rev.
  Condens. Matter Phys.}\ }\textbf {\bibinfo {volume} {1}},\ \bibinfo {pages}
  {71--88} (\bibinfo {year} {2010})}\BibitemShut {NoStop}%
\bibitem [{Spi(2012)}]{Spintronics_InsightNatureMater11/2012}%
  \BibitemOpen
  \href@noop {} {\enquote {\bibinfo {title} {Spintronics},}\ }\bibinfo
  {howpublished} {Insight issue of Nat. Mater. \textbf{11}, 367--416} (\bibinfo
  {year} {2012})\BibitemShut {NoStop}%
\bibitem [{\citenamefont {Hueso}\ \emph {et~al.}(2007)\citenamefont {Hueso},
  \citenamefont {Pruneda}, \citenamefont {Ferrari}, \citenamefont {Burnell},
  \citenamefont {Vald\'es-Herrera}, \citenamefont {Simons}, \citenamefont
  {Littlewood}, \citenamefont {Artacho}, \citenamefont {Fert},\ and\
  \citenamefont {Mathur}}]{Hueso07}%
  \BibitemOpen
  \bibfield  {author} {\bibinfo {author} {\bibfnamefont {L.~E.}\ \bibnamefont
  {Hueso}}, \bibinfo {author} {\bibfnamefont {J.~M.}\ \bibnamefont {Pruneda}},
  \bibinfo {author} {\bibfnamefont {V.}~\bibnamefont {Ferrari}}, \bibinfo
  {author} {\bibfnamefont {G.}~\bibnamefont {Burnell}}, \bibinfo {author}
  {\bibfnamefont {J.~P.}\ \bibnamefont {Vald\'es-Herrera}}, \bibinfo {author}
  {\bibfnamefont {B~.D.}\ \bibnamefont {Simons}}, \bibinfo {author}
  {\bibfnamefont {P.~B.}\ \bibnamefont {Littlewood}}, \bibinfo {author}
  {\bibfnamefont {E.}~\bibnamefont {Artacho}}, \bibinfo {author} {\bibfnamefont
  {A.}~\bibnamefont {Fert}}, \ and\ \bibinfo {author} {\bibfnamefont {N.~D.}\
  \bibnamefont {Mathur}},\ }\bibfield  {title} {\enquote {\bibinfo {title}
  {Transformation of spin information into large elctgrical signals using
  carbon nanotubes},}\ }\href@noop {} {\bibfield  {journal} {\bibinfo
  {journal} {Nature}\ }\textbf {\bibinfo {volume} {445}},\ \bibinfo {pages}
  {410--413} (\bibinfo {year} {2007})}\BibitemShut {NoStop}%
\bibitem [{\citenamefont {Dlubak}\ \emph {et~al.}(2012)\citenamefont {Dlubak},
  \citenamefont {Martin}, \citenamefont {Deranlot}, \citenamefont {Servet},
  \citenamefont {Xavier}, \citenamefont {Mattana}, \citenamefont {Sprinkle},
  \citenamefont {Berger}, \citenamefont {DeHeer}, \citenamefont {Petroff},
  \citenamefont {Anane}, \citenamefont {Seneor},\ and\ \citenamefont
  {Fert}}]{Dlubak12}%
  \BibitemOpen
  \bibfield  {author} {\bibinfo {author} {\bibfnamefont {B.}~\bibnamefont
  {Dlubak}}, \bibinfo {author} {\bibfnamefont {M-B.}\ \bibnamefont {Martin}},
  \bibinfo {author} {\bibfnamefont {C.}~\bibnamefont {Deranlot}}, \bibinfo
  {author} {\bibfnamefont {B.}~\bibnamefont {Servet}}, \bibinfo {author}
  {\bibfnamefont {S.}~\bibnamefont {Xavier}}, \bibinfo {author} {\bibfnamefont
  {R.}~\bibnamefont {Mattana}}, \bibinfo {author} {\bibfnamefont
  {M.}~\bibnamefont {Sprinkle}}, \bibinfo {author} {\bibfnamefont
  {C.}~\bibnamefont {Berger}}, \bibinfo {author} {\bibfnamefont {W.~A.}\
  \bibnamefont {DeHeer}}, \bibinfo {author} {\bibfnamefont {F.}~\bibnamefont
  {Petroff}}, \bibinfo {author} {\bibfnamefont {A.}~\bibnamefont {Anane}},
  \bibinfo {author} {\bibfnamefont {P.}~\bibnamefont {Seneor}}, \ and\ \bibinfo
  {author} {\bibfnamefont {A.}~\bibnamefont {Fert}},\ }\bibfield  {title}
  {\enquote {\bibinfo {title} {Highly efficient spin transport in epitaxial
  graphene on sic},}\ }\href@noop {} {\bibfield  {journal} {\bibinfo  {journal}
  {Nat. Phys.}\ }\textbf {\bibinfo {volume} {8}},\ \bibinfo {pages} {557--561}
  (\bibinfo {year} {2012})}\BibitemShut {NoStop}%
\bibitem [{\citenamefont {Han}\ \emph {et~al.}({2014})\citenamefont {Han},
  \citenamefont {Kawakami}, \citenamefont {Gmitra},\ and\ \citenamefont
  {Fabian}}]{Han14}%
  \BibitemOpen
  \bibfield  {author} {\bibinfo {author} {\bibfnamefont {W.}~\bibnamefont
  {Han}}, \bibinfo {author} {\bibfnamefont {R.~K.}\ \bibnamefont {Kawakami}},
  \bibinfo {author} {\bibfnamefont {M.}~\bibnamefont {Gmitra}}, \ and\ \bibinfo
  {author} {\bibfnamefont {J.}~\bibnamefont {Fabian}},\ }\bibfield  {title}
  {\enquote {\bibinfo {title} {{Graphene spintronics}},}\ }\href@noop {}
  {\bibfield  {journal} {\bibinfo  {journal} {{Nat. Nanotechnol.}}\ }\textbf
  {\bibinfo {volume} {{9}}},\ \bibinfo {pages} {{794--807}} (\bibinfo {year}
  {{2014}})}\BibitemShut {NoStop}%
\bibitem [{\citenamefont {Guimar\~aes}\ \emph {et~al.}(2014)\citenamefont
  {Guimar\~aes}, \citenamefont {Zomer}, \citenamefont {Ingla-Ayn\'es},
  \citenamefont {Brant}, \citenamefont {Tombros},\ and\ \citenamefont {van
  Wees}}]{guim14}%
  \BibitemOpen
  \bibfield  {author} {\bibinfo {author} {\bibfnamefont {M.H.D.}\ \bibnamefont
  {Guimar\~aes}}, \bibinfo {author} {\bibfnamefont {P.J.}\ \bibnamefont
  {Zomer}}, \bibinfo {author} {\bibfnamefont {J.}~\bibnamefont
  {Ingla-Ayn\'es}}, \bibinfo {author} {\bibfnamefont {J.C.}\ \bibnamefont
  {Brant}}, \bibinfo {author} {\bibfnamefont {N.}~\bibnamefont {Tombros}}, \
  and\ \bibinfo {author} {\bibfnamefont {B.J.}\ \bibnamefont {van Wees}},\
  }\bibfield  {title} {\enquote {\bibinfo {title} {Controlling spin relaxation
  in hexagonal bn-encapsulated graphene with a transverse electric field},}\
  }\href@noop {} {\bibfield  {journal} {\bibinfo  {journal} {Phys. Rev. Lett.}\
  }\textbf {\bibinfo {volume} {113}},\ \bibinfo {pages} {086602} (\bibinfo
  {year} {2014})}\BibitemShut {NoStop}%
\bibitem [{\citenamefont {Dr\"ogeler}\ \emph {et~al.}(2014)\citenamefont
  {Dr\"ogeler}, \citenamefont {Volmer}, \citenamefont {Wolter}, \citenamefont
  {Terr\'es}, \citenamefont {Watanabe}, \citenamefont {Taniguchi},
  \citenamefont {G\"untherodt}, \citenamefont {Stampfer},\ and\ \citenamefont
  {Beschoten}}]{Beschoten14}%
  \BibitemOpen
  \bibfield  {author} {\bibinfo {author} {\bibfnamefont {M.}~\bibnamefont
  {Dr\"ogeler}}, \bibinfo {author} {\bibfnamefont {F.}~\bibnamefont {Volmer}},
  \bibinfo {author} {\bibfnamefont {M.}~\bibnamefont {Wolter}}, \bibinfo
  {author} {\bibfnamefont {B.}~\bibnamefont {Terr\'es}}, \bibinfo {author}
  {\bibfnamefont {K.}~\bibnamefont {Watanabe}}, \bibinfo {author}
  {\bibfnamefont {T.}~\bibnamefont {Taniguchi}}, \bibinfo {author}
  {\bibfnamefont {G.}~\bibnamefont {G\"untherodt}}, \bibinfo {author}
  {\bibfnamefont {C.}~\bibnamefont {Stampfer}}, \ and\ \bibinfo {author}
  {\bibfnamefont {B.}~\bibnamefont {Beschoten}},\ }\bibfield  {title} {\enquote
  {\bibinfo {title} {{Nanosecond spin lifetimes in single-and few-layer
  graphene--hBN heterostructures at room temperature}},}\ }\href@noop {}
  {\bibfield  {journal} {\bibinfo  {journal} {Nano Lett.}\ }\textbf {\bibinfo
  {volume} {14}},\ \bibinfo {pages} {6050--6055} (\bibinfo {year}
  {2014})}\BibitemShut {NoStop}%
\bibitem [{\citenamefont {Laird}\ \emph {et~al.}({2013})\citenamefont {Laird},
  \citenamefont {Pei},\ and\ \citenamefont {Kouwenhoven}}]{Laird13}%
  \BibitemOpen
  \bibfield  {author} {\bibinfo {author} {\bibfnamefont {E.~A.}\ \bibnamefont
  {Laird}}, \bibinfo {author} {\bibfnamefont {F.}~\bibnamefont {Pei}}, \ and\
  \bibinfo {author} {\bibfnamefont {L.~P.}\ \bibnamefont {Kouwenhoven}},\
  }\bibfield  {title} {\enquote {\bibinfo {title} {{A valley-spin qubit in a
  carbon nanotube}},}\ }\href@noop {} {\bibfield  {journal} {\bibinfo
  {journal} {{Nat. Nanotechnol.}}\ }\textbf {\bibinfo {volume} {{8}}},\
  \bibinfo {pages} {{565--568}} (\bibinfo {year} {{2013}})}\BibitemShut
  {NoStop}%
\bibitem [{\citenamefont {Viennot}\ \emph {et~al.}({2015})\citenamefont
  {Viennot}, \citenamefont {Dartiailh}, \citenamefont {Cottet},\ and\
  \citenamefont {Kontos}}]{Viennot15}%
  \BibitemOpen
  \bibfield  {author} {\bibinfo {author} {\bibfnamefont {J.~J.}\ \bibnamefont
  {Viennot}}, \bibinfo {author} {\bibfnamefont {M.~C.}\ \bibnamefont
  {Dartiailh}}, \bibinfo {author} {\bibfnamefont {A.}~\bibnamefont {Cottet}}, \
  and\ \bibinfo {author} {\bibfnamefont {T.}~\bibnamefont {Kontos}},\
  }\bibfield  {title} {\enquote {\bibinfo {title} {{Coherent coupling of a
  single spin to microwave cavity photons}},}\ }\href@noop {} {\bibfield
  {journal} {\bibinfo  {journal} {{Science}}\ }\textbf {\bibinfo {volume}
  {{349}}},\ \bibinfo {pages} {{408--411}} (\bibinfo {year}
  {{2015}})}\BibitemShut {NoStop}%
\bibitem [{\citenamefont {Morgan}\ \emph {et~al.}(2016)\citenamefont {Morgan},
  \citenamefont {Misiorny}, \citenamefont {Metten}, \citenamefont {Heedt},
  \citenamefont {Sch\"appears}, \citenamefont {Schneider},\ and\ \citenamefont
  {Meyer}}]{Morgan16}%
  \BibitemOpen
  \bibfield  {author} {\bibinfo {author} {\bibfnamefont {C.}~\bibnamefont
  {Morgan}}, \bibinfo {author} {\bibfnamefont {M.}~\bibnamefont {Misiorny}},
  \bibinfo {author} {\bibfnamefont {D.}~\bibnamefont {Metten}}, \bibinfo
  {author} {\bibfnamefont {S.}~\bibnamefont {Heedt}}, \bibinfo {author}
  {\bibfnamefont {Th.}\ \bibnamefont {Sch\"appears}}, \bibinfo {author}
  {\bibfnamefont {C.~M.}\ \bibnamefont {Schneider}}, \ and\ \bibinfo {author}
  {\bibfnamefont {C.}~\bibnamefont {Meyer}},\ }\bibfield  {title} {\enquote
  {\bibinfo {title} {Impact of tunnel barrier strength on magnetoresistance in
  carbon nanotubes},}\ }\href@noop {} {\bibfield  {journal} {\bibinfo
  {journal} {Phys. Rev. Appl.}\ }\textbf {\bibinfo {volume} {5}},\ \bibinfo
  {pages} {054010} (\bibinfo {year} {2016})}\BibitemShut {NoStop}%
\bibitem [{\citenamefont {Han}\ \emph {et~al.}(2010)\citenamefont {Han},
  \citenamefont {Pi}, \citenamefont {McCreary}, \citenamefont {Li},
  \citenamefont {Wong}, \citenamefont {Swartz},\ and\ \citenamefont
  {Kawakami}}]{Han10}%
  \BibitemOpen
  \bibfield  {author} {\bibinfo {author} {\bibfnamefont {W.}~\bibnamefont
  {Han}}, \bibinfo {author} {\bibfnamefont {K.}~\bibnamefont {Pi}}, \bibinfo
  {author} {\bibfnamefont {K.~M.}\ \bibnamefont {McCreary}}, \bibinfo {author}
  {\bibfnamefont {Yan}\ \bibnamefont {Li}}, \bibinfo {author} {\bibfnamefont
  {Jared J.~I.}\ \bibnamefont {Wong}}, \bibinfo {author} {\bibfnamefont
  {A.~G.}\ \bibnamefont {Swartz}}, \ and\ \bibinfo {author} {\bibfnamefont
  {R.~K.}\ \bibnamefont {Kawakami}},\ }\bibfield  {title} {\enquote {\bibinfo
  {title} {Tunneling spin injection into single layer graphene},}\ }\href@noop
  {} {\bibfield  {journal} {\bibinfo  {journal} {Phys. Rev. Lett.}\ }\textbf
  {\bibinfo {volume} {105}},\ \bibinfo {pages} {167202} (\bibinfo {year}
  {2010})}\BibitemShut {NoStop}%
\bibitem [{\citenamefont {\ifmmode \check{Z}\else
  \v{Z}\fi{}uti\ifmmode~\acute{c}\else \'{c}\fi{}}\ \emph
  {et~al.}(2004)\citenamefont {\ifmmode \check{Z}\else
  \v{Z}\fi{}uti\ifmmode~\acute{c}\else \'{c}\fi{}}, \citenamefont {Fabian},\
  and\ \citenamefont {Das~Sarma}}]{Zutic04}%
  \BibitemOpen
  \bibfield  {author} {\bibinfo {author} {\bibfnamefont {I.}~\bibnamefont
  {\ifmmode \check{Z}\else \v{Z}\fi{}uti\ifmmode~\acute{c}\else \'{c}\fi{}}},
  \bibinfo {author} {\bibfnamefont {J.}~\bibnamefont {Fabian}}, \ and\ \bibinfo
  {author} {\bibfnamefont {S.}~\bibnamefont {Das~Sarma}},\ }\bibfield  {title}
  {\enquote {\bibinfo {title} {Spintronics: Fundamentals and applications},}\
  }\href@noop {} {\bibfield  {journal} {\bibinfo  {journal} {Rev. Mod. Phys.}\
  }\textbf {\bibinfo {volume} {76}},\ \bibinfo {pages} {323--410} (\bibinfo
  {year} {2004})}\BibitemShut {NoStop}%
\bibitem [{\citenamefont {Sinova}\ \emph {et~al.}(2015)\citenamefont {Sinova},
  \citenamefont {Valenzuela}, \citenamefont {Wunderlich}, \citenamefont
  {Back},\ and\ \citenamefont {Jungwirth}}]{Sinova_Rev.Mod.Phys.87/2015}%
  \BibitemOpen
  \bibfield  {author} {\bibinfo {author} {\bibfnamefont {J}~\bibnamefont
  {Sinova}}, \bibinfo {author} {\bibfnamefont {S}~\bibnamefont {Valenzuela}},
  \bibinfo {author} {\bibfnamefont {J.}~\bibnamefont {Wunderlich}}, \bibinfo
  {author} {\bibfnamefont {C.~H.}\ \bibnamefont {Back}}, \ and\ \bibinfo
  {author} {\bibfnamefont {T.}~\bibnamefont {Jungwirth}},\ }\bibfield  {title}
  {\enquote {\bibinfo {title} {{Spin Hall effects}},}\ }\href@noop {}
  {\bibfield  {journal} {\bibinfo  {journal} {Rev. Mod. Phys.}\ }\textbf
  {\bibinfo {volume} {87}},\ \bibinfo {pages} {1213--1259} (\bibinfo {year}
  {2015})}\BibitemShut {NoStop}%
\bibitem [{\citenamefont {Gmitra}\ \emph {et~al.}(2009)\citenamefont {Gmitra},
  \citenamefont {Konschuh}, \citenamefont {Ertler}, \citenamefont
  {Ambrosch-Draxl},\ and\ \citenamefont {Fabian}}]{Gmitra09}%
  \BibitemOpen
  \bibfield  {author} {\bibinfo {author} {\bibfnamefont {M.}~\bibnamefont
  {Gmitra}}, \bibinfo {author} {\bibfnamefont {S.}~\bibnamefont {Konschuh}},
  \bibinfo {author} {\bibfnamefont {C.}~\bibnamefont {Ertler}}, \bibinfo
  {author} {\bibfnamefont {C.}~\bibnamefont {Ambrosch-Draxl}}, \ and\ \bibinfo
  {author} {\bibfnamefont {J.}~\bibnamefont {Fabian}},\ }\bibfield  {title}
  {\enquote {\bibinfo {title} {Band-structure topologies of graphene:
  Spin-orbit coupling effects from first principles},}\ }\href@noop {}
  {\bibfield  {journal} {\bibinfo  {journal} {Phys. Rev. B}\ }\textbf {\bibinfo
  {volume} {80}},\ \bibinfo {pages} {235431} (\bibinfo {year}
  {2009})}\BibitemShut {NoStop}%
\bibitem [{\citenamefont {Castro~Neto}\ and\ \citenamefont
  {Guinea}(2009)}]{Castro09}%
  \BibitemOpen
  \bibfield  {author} {\bibinfo {author} {\bibfnamefont {A.~H.}\ \bibnamefont
  {Castro~Neto}}\ and\ \bibinfo {author} {\bibfnamefont {F.}~\bibnamefont
  {Guinea}},\ }\bibfield  {title} {\enquote {\bibinfo {title} {Impurity-induced
  spin-orbit coupling in graphene},}\ }\href@noop {} {\bibfield  {journal}
  {\bibinfo  {journal} {Phys. Rev. Lett.}\ }\textbf {\bibinfo {volume} {103}},\
  \bibinfo {pages} {026804} (\bibinfo {year} {2009})}\BibitemShut {NoStop}%
\bibitem [{\citenamefont {Huertas-Hernando}\ \emph {et~al.}(2006)\citenamefont
  {Huertas-Hernando}, \citenamefont {Guinea},\ and\ \citenamefont
  {Brataas}}]{Huertas06}%
  \BibitemOpen
  \bibfield  {author} {\bibinfo {author} {\bibfnamefont {D.}~\bibnamefont
  {Huertas-Hernando}}, \bibinfo {author} {\bibfnamefont {F.}~\bibnamefont
  {Guinea}}, \ and\ \bibinfo {author} {\bibfnamefont {A.}~\bibnamefont
  {Brataas}},\ }\bibfield  {title} {\enquote {\bibinfo {title} {Spin-orbit
  coupling in curved graphene, fullerenes, nanotubes, and nanotube caps},}\
  }\href@noop {} {\bibfield  {journal} {\bibinfo  {journal} {Phys. Rev. B}\
  }\textbf {\bibinfo {volume} {74}},\ \bibinfo {pages} {155426} (\bibinfo
  {year} {2006})}\BibitemShut {NoStop}%
\bibitem [{\citenamefont {Jhang}\ \emph {et~al.}(2010)\citenamefont {Jhang},
  \citenamefont {Marganska}, \citenamefont {Skourski}, \citenamefont
  {Preusche}, \citenamefont {Witkamp}, \citenamefont {Grifoni}, \citenamefont
  {van~der Zant}, \citenamefont {Wosnitza},\ and\ \citenamefont
  {Strunk}}]{Jhang10}%
  \BibitemOpen
  \bibfield  {author} {\bibinfo {author} {\bibfnamefont {S.~H.}\ \bibnamefont
  {Jhang}}, \bibinfo {author} {\bibfnamefont {M.}~\bibnamefont {Marganska}},
  \bibinfo {author} {\bibfnamefont {Y.}~\bibnamefont {Skourski}}, \bibinfo
  {author} {\bibfnamefont {D.}~\bibnamefont {Preusche}}, \bibinfo {author}
  {\bibfnamefont {B.}~\bibnamefont {Witkamp}}, \bibinfo {author} {\bibfnamefont
  {M.}~\bibnamefont {Grifoni}}, \bibinfo {author} {\bibfnamefont
  {H.}~\bibnamefont {van~der Zant}}, \bibinfo {author} {\bibfnamefont
  {J.}~\bibnamefont {Wosnitza}}, \ and\ \bibinfo {author} {\bibfnamefont
  {C.}~\bibnamefont {Strunk}},\ }\bibfield  {title} {\enquote {\bibinfo {title}
  {Spin-orbit interaction in chiral carbon nanotubes probed in pulsed magnetic
  fields},}\ }\href@noop {} {\bibfield  {journal} {\bibinfo  {journal} {Phys.
  Rev. B}\ }\textbf {\bibinfo {volume} {82}},\ \bibinfo {pages} {041404}
  (\bibinfo {year} {2010})}\BibitemShut {NoStop}%
\bibitem [{\citenamefont {Steele}\ \emph {et~al.}(2013)\citenamefont {Steele},
  \citenamefont {Pei}, \citenamefont {Laird}, \citenamefont {Jol},
  \citenamefont {Meerwaldt},\ and\ \citenamefont {Kouwenhoven}}]{Steele13}%
  \BibitemOpen
  \bibfield  {author} {\bibinfo {author} {\bibfnamefont {G.~A.}\ \bibnamefont
  {Steele}}, \bibinfo {author} {\bibfnamefont {F.}~\bibnamefont {Pei}},
  \bibinfo {author} {\bibfnamefont {E.~A.}\ \bibnamefont {Laird}}, \bibinfo
  {author} {\bibfnamefont {J.~M.}\ \bibnamefont {Jol}}, \bibinfo {author}
  {\bibfnamefont {H.~B.}\ \bibnamefont {Meerwaldt}}, \ and\ \bibinfo {author}
  {\bibfnamefont {L.~P.}\ \bibnamefont {Kouwenhoven}},\ }\bibfield  {title}
  {\enquote {\bibinfo {title} {{Large spin-orbit coupling in carbon
  nanotubes}},}\ }\href@noop {} {\bibfield  {journal} {\bibinfo  {journal}
  {Nat. Commun.}\ }\textbf {\bibinfo {volume} {4}},\ \bibinfo {pages} {1573}
  (\bibinfo {year} {2013})}\BibitemShut {NoStop}%
\bibitem [{\citenamefont {Tsukagoshi}\ \emph {et~al.}(1999)\citenamefont
  {Tsukagoshi}, \citenamefont {Alphenaar},\ and\ \citenamefont
  {Ago}}]{Tsukagoshi_Nature401/1999}%
  \BibitemOpen
  \bibfield  {author} {\bibinfo {author} {\bibfnamefont {K.}~\bibnamefont
  {Tsukagoshi}}, \bibinfo {author} {\bibfnamefont {B.~W.}\ \bibnamefont
  {Alphenaar}}, \ and\ \bibinfo {author} {\bibfnamefont {H.}~\bibnamefont
  {Ago}},\ }\bibfield  {title} {\enquote {\bibinfo {title} {Coherent transport
  of electron spin in a ferromagnetically contacted carbon nanotube},}\
  }\href@noop {} {\bibfield  {journal} {\bibinfo  {journal} {Nature}\ }\textbf
  {\bibinfo {volume} {401}},\ \bibinfo {pages} {572--574} (\bibinfo {year}
  {1999})}\BibitemShut {NoStop}%
\bibitem [{\citenamefont {Kim}\ \emph {et~al.}(2002)\citenamefont {Kim},
  \citenamefont {So}, \citenamefont {Kim},\ and\ \citenamefont
  {Kim}}]{Kim_Phys.Rev.B66/2002}%
  \BibitemOpen
  \bibfield  {author} {\bibinfo {author} {\bibfnamefont {J.-R.}\ \bibnamefont
  {Kim}}, \bibinfo {author} {\bibfnamefont {H.~Mi}\ \bibnamefont {So}},
  \bibinfo {author} {\bibfnamefont {J.-J.}\ \bibnamefont {Kim}}, \ and\
  \bibinfo {author} {\bibfnamefont {J.}~\bibnamefont {Kim}},\ }\bibfield
  {title} {\enquote {\bibinfo {title} {Spin-dependent transport properties in a
  single-walled carbon nanotube with mesoscopic co contacts},}\ }\href@noop {}
  {\bibfield  {journal} {\bibinfo  {journal} {Phys. Rev. B}\ }\textbf {\bibinfo
  {volume} {66}},\ \bibinfo {pages} {233401} (\bibinfo {year}
  {2002})}\BibitemShut {NoStop}%
\bibitem [{\citenamefont {Zhao}\ \emph {et~al.}(2002)\citenamefont {Zhao},
  \citenamefont {M{\"o}nch}, \citenamefont {Vinzelberg}, \citenamefont
  {M{\"u}hl},\ and\ \citenamefont {Schneider}}]{Zhao_Appl.Phys.Lett.80/2002}%
  \BibitemOpen
  \bibfield  {author} {\bibinfo {author} {\bibfnamefont {B.}~\bibnamefont
  {Zhao}}, \bibinfo {author} {\bibfnamefont {I.}~\bibnamefont {M{\"o}nch}},
  \bibinfo {author} {\bibfnamefont {H.}~\bibnamefont {Vinzelberg}}, \bibinfo
  {author} {\bibfnamefont {T.}~\bibnamefont {M{\"u}hl}}, \ and\ \bibinfo
  {author} {\bibfnamefont {C.M.}\ \bibnamefont {Schneider}},\ }\bibfield
  {title} {\enquote {\bibinfo {title} {Spin-coherent transport in
  ferromagnetically contacted carbon nanotubes},}\ }\href@noop {} {\bibfield
  {journal} {\bibinfo  {journal} {Appl. Phys. Lett.}\ }\textbf {\bibinfo
  {volume} {80}},\ \bibinfo {pages} {3144--3146} (\bibinfo {year}
  {2002})}\BibitemShut {NoStop}%
\bibitem [{\citenamefont {Jensen}\ \emph {et~al.}(2005)\citenamefont {Jensen},
  \citenamefont {Hauptmann}, \citenamefont {Nyg\aa{}rd},\ and\ \citenamefont
  {Lindelof}}]{Jensen_Phys.Rev.B72/2005}%
  \BibitemOpen
  \bibfield  {author} {\bibinfo {author} {\bibfnamefont {A.}~\bibnamefont
  {Jensen}}, \bibinfo {author} {\bibfnamefont {J.R.}\ \bibnamefont
  {Hauptmann}}, \bibinfo {author} {\bibfnamefont {J.}~\bibnamefont
  {Nyg\aa{}rd}}, \ and\ \bibinfo {author} {\bibfnamefont {P.E.}\ \bibnamefont
  {Lindelof}},\ }\bibfield  {title} {\enquote {\bibinfo {title}
  {Magnetoresistance in ferromagnetically contacted single-wall carbon
  nanotubes},}\ }\href@noop {} {\bibfield  {journal} {\bibinfo  {journal}
  {Phys. Rev. B}\ }\textbf {\bibinfo {volume} {72}},\ \bibinfo {pages} {035419}
  (\bibinfo {year} {2005})}\BibitemShut {NoStop}%
\bibitem [{\citenamefont {Sahoo}\ \emph
  {et~al.}(2005{\natexlab{a}})\citenamefont {Sahoo}, \citenamefont {Kontos},
  \citenamefont {Sch{\"o}nenberger},\ and\ \citenamefont
  {S{\"u}rgers}}]{Sahoo_Appl.Phys.Lett.86/2005}%
  \BibitemOpen
  \bibfield  {author} {\bibinfo {author} {\bibfnamefont {S.}~\bibnamefont
  {Sahoo}}, \bibinfo {author} {\bibfnamefont {T.}~\bibnamefont {Kontos}},
  \bibinfo {author} {\bibfnamefont {C.}~\bibnamefont {Sch{\"o}nenberger}}, \
  and\ \bibinfo {author} {\bibfnamefont {C.}~\bibnamefont {S{\"u}rgers}},\
  }\bibfield  {title} {\enquote {\bibinfo {title} {Electrical spin injection in
  multiwall carbon nanotubes with transparent ferromagnetic contacts},}\
  }\href@noop {} {\bibfield  {journal} {\bibinfo  {journal} {Appl. Phys.
  Lett.}\ }\textbf {\bibinfo {volume} {86}} (\bibinfo {year}
  {2005}{\natexlab{a}})}\BibitemShut {NoStop}%
\bibitem [{\citenamefont {Man}\ \emph {et~al.}(2006)\citenamefont {Man},
  \citenamefont {Wever},\ and\ \citenamefont
  {Morpurgo}}]{Man_Phys.Rev.B73/2006}%
  \BibitemOpen
  \bibfield  {author} {\bibinfo {author} {\bibfnamefont {H.~T.}\ \bibnamefont
  {Man}}, \bibinfo {author} {\bibfnamefont {I.~J.~W.}\ \bibnamefont {Wever}}, \
  and\ \bibinfo {author} {\bibfnamefont {A.~F.}\ \bibnamefont {Morpurgo}},\
  }\bibfield  {title} {\enquote {\bibinfo {title} {Spin-dependent quantum
  interference in single-wall carbon nanotubes with ferromagnetic contacts},}\
  }\href@noop {} {\bibfield  {journal} {\bibinfo  {journal} {Phys. Rev. B}\
  }\textbf {\bibinfo {volume} {73}},\ \bibinfo {pages} {241401} (\bibinfo
  {year} {2006})}\BibitemShut {NoStop}%
\bibitem [{\citenamefont {Liang}\ \emph {et~al.}(2001)\citenamefont {Liang},
  \citenamefont {Bockrath}, \citenamefont {Bozovic}, \citenamefont {Hafner},
  \citenamefont {Tinkham},\ and\ \citenamefont {Park}}]{Liang_Nature411/2001}%
  \BibitemOpen
  \bibfield  {author} {\bibinfo {author} {\bibfnamefont {W.}~\bibnamefont
  {Liang}}, \bibinfo {author} {\bibfnamefont {M.}~\bibnamefont {Bockrath}},
  \bibinfo {author} {\bibfnamefont {D.}~\bibnamefont {Bozovic}}, \bibinfo
  {author} {\bibfnamefont {J.H.}\ \bibnamefont {Hafner}}, \bibinfo {author}
  {\bibfnamefont {M.}~\bibnamefont {Tinkham}}, \ and\ \bibinfo {author}
  {\bibfnamefont {H.}~\bibnamefont {Park}},\ }\bibfield  {title} {\enquote
  {\bibinfo {title} {{Fabry-Perot interference in a nanotube electron
  waveguide}},}\ }\href@noop {} {\bibfield  {journal} {\bibinfo  {journal}
  {Nature}\ }\textbf {\bibinfo {volume} {411}},\ \bibinfo {pages} {665--669}
  (\bibinfo {year} {2001})}\BibitemShut {NoStop}%
\bibitem [{\citenamefont {Sahoo}\ \emph
  {et~al.}(2005{\natexlab{b}})\citenamefont {Sahoo}, \citenamefont {Kontos},
  \citenamefont {Furer}, \citenamefont {Hoffmann}, \citenamefont {Gr{\"a}ber},
  \citenamefont {Cottet},\ and\ \citenamefont
  {Sch{\"o}nenberger}}]{Sahoo_NaturePhys.1/2005}%
  \BibitemOpen
  \bibfield  {author} {\bibinfo {author} {\bibfnamefont {S.}~\bibnamefont
  {Sahoo}}, \bibinfo {author} {\bibfnamefont {T.}~\bibnamefont {Kontos}},
  \bibinfo {author} {\bibfnamefont {J.}~\bibnamefont {Furer}}, \bibinfo
  {author} {\bibfnamefont {C.}~\bibnamefont {Hoffmann}}, \bibinfo {author}
  {\bibfnamefont {M.}~\bibnamefont {Gr{\"a}ber}}, \bibinfo {author}
  {\bibfnamefont {A.}~\bibnamefont {Cottet}}, \ and\ \bibinfo {author}
  {\bibfnamefont {C.}~\bibnamefont {Sch{\"o}nenberger}},\ }\bibfield  {title}
  {\enquote {\bibinfo {title} {Electric field control of spin transport},}\
  }\href@noop {} {\bibfield  {journal} {\bibinfo  {journal} {Nat. Phys.}\
  }\textbf {\bibinfo {volume} {1}},\ \bibinfo {pages} {99--102} (\bibinfo
  {year} {2005}{\natexlab{b}})}\BibitemShut {NoStop}%
\bibitem [{\citenamefont {Man}\ and\ \citenamefont
  {Morpurgo}(2005)}]{Man_Phys.Rev.Lett.95/2005}%
  \BibitemOpen
  \bibfield  {author} {\bibinfo {author} {\bibfnamefont {H.T.}\ \bibnamefont
  {Man}}\ and\ \bibinfo {author} {\bibfnamefont {A.F.}\ \bibnamefont
  {Morpurgo}},\ }\bibfield  {title} {\enquote {\bibinfo {title}
  {Sample-specific and ensemble-averaged magnetoconductance of individual
  single-wall carbon nanotubes},}\ }\href@noop {} {\bibfield  {journal}
  {\bibinfo  {journal} {Phys. Rev. Lett.}\ }\textbf {\bibinfo {volume} {95}},\
  \bibinfo {pages} {026801} (\bibinfo {year} {2005})}\BibitemShut {NoStop}%
\bibitem [{\citenamefont {Cottet}\ \emph
  {et~al.}(2006{\natexlab{a}})\citenamefont {Cottet}, \citenamefont {Kontos},
  \citenamefont {Belzig}, \citenamefont {Sch\"{o}nenberger},\ and\
  \citenamefont {Bruder}}]{Cottet_Europhys.Lett.74/2006}%
  \BibitemOpen
  \bibfield  {author} {\bibinfo {author} {\bibfnamefont {A.}~\bibnamefont
  {Cottet}}, \bibinfo {author} {\bibfnamefont {T.}~\bibnamefont {Kontos}},
  \bibinfo {author} {\bibfnamefont {W.}~\bibnamefont {Belzig}}, \bibinfo
  {author} {\bibfnamefont {C.}~\bibnamefont {Sch\"{o}nenberger}}, \ and\
  \bibinfo {author} {\bibfnamefont {C.}~\bibnamefont {Bruder}},\ }\bibfield
  {title} {\enquote {\bibinfo {title} {Controlling spin in an electronic
  interferometer with spin-active interfaces},}\ }\href@noop {} {\bibfield
  {journal} {\bibinfo  {journal} {Europhys. Lett.}\ }\textbf {\bibinfo {volume}
  {74}},\ \bibinfo {pages} {320--326} (\bibinfo {year}
  {2006}{\natexlab{a}})}\BibitemShut {NoStop}%
\bibitem [{\citenamefont {Cottet}\ \emph
  {et~al.}(2006{\natexlab{b}})\citenamefont {Cottet}, \citenamefont {Kontos},
  \citenamefont {Sahoo}, \citenamefont {Man}, \citenamefont {Choi},
  \citenamefont {Belzig}, \citenamefont {Bruder}, \citenamefont {Morpurgo},\
  and\ \citenamefont {Sch{\"o}nenberger}}]{Cottet_Sem.Sci.Tech.21/2006}%
  \BibitemOpen
  \bibfield  {author} {\bibinfo {author} {\bibfnamefont {A.}~\bibnamefont
  {Cottet}}, \bibinfo {author} {\bibfnamefont {T.}~\bibnamefont {Kontos}},
  \bibinfo {author} {\bibfnamefont {S.}~\bibnamefont {Sahoo}}, \bibinfo
  {author} {\bibfnamefont {H.T.}\ \bibnamefont {Man}}, \bibinfo {author}
  {\bibfnamefont {M.-S.}\ \bibnamefont {Choi}}, \bibinfo {author}
  {\bibfnamefont {W.}~\bibnamefont {Belzig}}, \bibinfo {author} {\bibfnamefont
  {C.}~\bibnamefont {Bruder}}, \bibinfo {author} {\bibfnamefont {A.F.}\
  \bibnamefont {Morpurgo}}, \ and\ \bibinfo {author} {\bibfnamefont
  {C.}~\bibnamefont {Sch{\"o}nenberger}},\ }\bibfield  {title} {\enquote
  {\bibinfo {title} {Nanospintronics with carbon nanotubes},}\ }\href@noop {}
  {\bibfield  {journal} {\bibinfo  {journal} {Sem. Sci. Tech.}\ }\textbf
  {\bibinfo {volume} {21}},\ \bibinfo {pages} {S78--S95} (\bibinfo {year}
  {2006}{\natexlab{b}})}\BibitemShut {NoStop}%
\bibitem [{\citenamefont {Grundler}(2001)}]{Grundler_Phys.Rev.Lett.86/2001}%
  \BibitemOpen
  \bibfield  {author} {\bibinfo {author} {\bibfnamefont {D.}~\bibnamefont
  {Grundler}},\ }\bibfield  {title} {\enquote {\bibinfo {title} {Oscillatory
  spin-filtering due to gate control of spin-dependent interface
  conductance},}\ }\href@noop {} {\bibfield  {journal} {\bibinfo  {journal}
  {Phys. Rev. Lett.}\ }\textbf {\bibinfo {volume} {86}},\ \bibinfo {pages}
  {1058--1061} (\bibinfo {year} {2001})}\BibitemShut {NoStop}%
\bibitem [{\citenamefont {Hu}\ and\ \citenamefont
  {Matsuyama}(2001)}]{Hu_Phys.Rev.Lett.87/2001}%
  \BibitemOpen
  \bibfield  {author} {\bibinfo {author} {\bibfnamefont {C.-M.}\ \bibnamefont
  {Hu}}\ and\ \bibinfo {author} {\bibfnamefont {T.}~\bibnamefont {Matsuyama}},\
  }\bibfield  {title} {\enquote {\bibinfo {title} {Spin injection across a
  heterojunction: A ballistic picture},}\ }\href@noop {} {\bibfield  {journal}
  {\bibinfo  {journal} {Phys. Rev. Lett.}\ }\textbf {\bibinfo {volume} {87}},\
  \bibinfo {pages} {066803} (\bibinfo {year} {2001})}\BibitemShut {NoStop}%
\bibitem [{\citenamefont {Kane}\ and\ \citenamefont
  {Mele}(1997)}]{Kane_Phys.Rev.Lett.78/1997}%
  \BibitemOpen
  \bibfield  {author} {\bibinfo {author} {\bibfnamefont {C.L.}\ \bibnamefont
  {Kane}}\ and\ \bibinfo {author} {\bibfnamefont {E.J.}\ \bibnamefont {Mele}},\
  }\bibfield  {title} {\enquote {\bibinfo {title} {Size, shape, and low energy
  electronic structure of carbon nanotubes},}\ }\href@noop {} {\bibfield
  {journal} {\bibinfo  {journal} {Phys. Rev. Lett.}\ }\textbf {\bibinfo
  {volume} {78}},\ \bibinfo {pages} {1932--1935} (\bibinfo {year}
  {1997})}\BibitemShut {NoStop}%
\bibitem [{\citenamefont {Egger}\ and\ \citenamefont
  {Gogolin}(1998)}]{Egger_Eur.Phys.J.B3/1998}%
  \BibitemOpen
  \bibfield  {author} {\bibinfo {author} {\bibfnamefont {R.}~\bibnamefont
  {Egger}}\ and\ \bibinfo {author} {\bibfnamefont {A.O.}\ \bibnamefont
  {Gogolin}},\ }\bibfield  {title} {\enquote {\bibinfo {title} {Correlated
  transport and non-fermi-liquid behavior in single-wall carbon nanotubes},}\
  }\href@noop {} {\bibfield  {journal} {\bibinfo  {journal} {Eur. Phys. J. B}\
  }\textbf {\bibinfo {volume} {3}},\ \bibinfo {pages} {281--300} (\bibinfo
  {year} {1998})}\BibitemShut {NoStop}%
\bibitem [{\citenamefont {Blonder}\ \emph {et~al.}(1982)\citenamefont
  {Blonder}, \citenamefont {Tinkham},\ and\ \citenamefont
  {Klapwijk}}]{Blonder_Phys.Rev.B25/1982}%
  \BibitemOpen
  \bibfield  {author} {\bibinfo {author} {\bibfnamefont {G.E.}\ \bibnamefont
  {Blonder}}, \bibinfo {author} {\bibfnamefont {M.}~\bibnamefont {Tinkham}}, \
  and\ \bibinfo {author} {\bibfnamefont {T.M.}\ \bibnamefont {Klapwijk}},\
  }\bibfield  {title} {\enquote {\bibinfo {title} {Transition from metallic to
  tunneling regimes in superconducting microconstrictions: Excess current,
  charge imbalance, and supercurrent conversion},}\ }\href@noop {} {\bibfield
  {journal} {\bibinfo  {journal} {Phys. Rev. B}\ }\textbf {\bibinfo {volume}
  {25}},\ \bibinfo {pages} {4515--4532} (\bibinfo {year} {1982})}\BibitemShut
  {NoStop}%
\bibitem [{\citenamefont {Qi}\ \emph {et~al.}(1998)\citenamefont {Qi},
  \citenamefont {Xing},\ and\ \citenamefont {Dong}}]{Qi_Phys.Rev.B58/1998}%
  \BibitemOpen
  \bibfield  {author} {\bibinfo {author} {\bibfnamefont {Y.}~\bibnamefont
  {Qi}}, \bibinfo {author} {\bibfnamefont {D.Y.}\ \bibnamefont {Xing}}, \ and\
  \bibinfo {author} {\bibfnamefont {J.}~\bibnamefont {Dong}},\ }\bibfield
  {title} {\enquote {\bibinfo {title} {Relation between julliere and
  slonczewski models of tunneling magnetoresistance},}\ }\href@noop {}
  {\bibfield  {journal} {\bibinfo  {journal} {Phys. Rev. B}\ }\textbf {\bibinfo
  {volume} {58}},\ \bibinfo {pages} {2783--2787} (\bibinfo {year}
  {1998})}\BibitemShut {NoStop}%
\bibitem [{\citenamefont {Maekawa}\ \emph {et~al.}(2012)\citenamefont
  {Maekawa}, \citenamefont {Valenzuela}, \citenamefont {Saitoh},\ and\
  \citenamefont {Kimura}}]{Maekawa_book}%
  \BibitemOpen
  \bibinfo {editor} {\bibfnamefont {S.}~\bibnamefont {Maekawa}}, \bibinfo
  {editor} {\bibfnamefont {S.}~\bibnamefont {Valenzuela}}, \bibinfo {editor}
  {\bibfnamefont {E.}~\bibnamefont {Saitoh}}, \ and\ \bibinfo {editor}
  {\bibfnamefont {T.}~\bibnamefont {Kimura}},\ eds.,\ \href@noop {} {\emph
  {\bibinfo {title} {Spin current}}},\ \bibinfo {series} {Series on
  Semiconductor Science and Technology}, Vol.~\bibinfo {volume} {17}\ (\bibinfo
   {publisher} {Oxford Univeristy Press},\ \bibinfo {address} {Oxford},\
  \bibinfo {year} {2012})\BibitemShut {NoStop}%
\bibitem [{\citenamefont {Tsymbal}\ and\ \citenamefont
  {\v{Z}uti\'{c}}(2012)}]{Tsybmal_book}%
  \BibitemOpen
  \bibinfo {editor} {\bibfnamefont {E.~Y.}\ \bibnamefont {Tsymbal}}\ and\
  \bibinfo {editor} {\bibfnamefont {I.}~\bibnamefont {\v{Z}uti\'{c}}},\ eds.,\
  \href@noop {} {\emph {\bibinfo {title} {Handbook of spin transport and
  magnetism}}}\ (\bibinfo  {publisher} {CRC Press},\ \bibinfo {address} {Boca
  Raton},\ \bibinfo {year} {2012})\BibitemShut {NoStop}%
\bibitem [{\citenamefont {De~Martino}\ and\ \citenamefont
  {Egger}(2005)}]{DeMartino_J.Phys.:Condens.Matter17/2005}%
  \BibitemOpen
  \bibfield  {author} {\bibinfo {author} {\bibfnamefont {A.}~\bibnamefont
  {De~Martino}}\ and\ \bibinfo {author} {\bibfnamefont {R.}~\bibnamefont
  {Egger}},\ }\bibfield  {title} {\enquote {\bibinfo {title} {Rashba
  spin--orbit coupling and spin precession in carbon nanotubes},}\ }\href@noop
  {} {\bibfield  {journal} {\bibinfo  {journal} {J. Phys.: Condens. Matter}\
  }\textbf {\bibinfo {volume} {17}},\ \bibinfo {pages} {5523--5532} (\bibinfo
  {year} {2005})}\BibitemShut {NoStop}%
\bibitem [{\citenamefont {Giamarchi}(2003)}]{Giamarchi_book}%
  \BibitemOpen
  \bibfield  {author} {\bibinfo {author} {\bibfnamefont {T.}~\bibnamefont
  {Giamarchi}},\ }\href@noop {} {\emph {\bibinfo {title} {Quantum physics in
  one dimension}}},\ \bibinfo {series} {International Series of Monographs on
  Physics}, Vol.\ \bibinfo {volume} {121}\ (\bibinfo  {publisher} {OUP},\
  \bibinfo {address} {Oxford},\ \bibinfo {year} {2003})\BibitemShut {NoStop}%
\bibitem [{\citenamefont {Kr{\"u}ger}\ \emph {et~al.}(2001)\citenamefont
  {Kr{\"u}ger}, \citenamefont {Buitelaar}, \citenamefont {Nussbaumer},
  \citenamefont {Sch{\"o}nenberger},\ and\ \citenamefont
  {Forr\'{o}}}]{Krueger_App.Phys.Lett.78/2001}%
  \BibitemOpen
  \bibfield  {author} {\bibinfo {author} {\bibfnamefont {M.}~\bibnamefont
  {Kr{\"u}ger}}, \bibinfo {author} {\bibfnamefont {M.R.}\ \bibnamefont
  {Buitelaar}}, \bibinfo {author} {\bibfnamefont {T.}~\bibnamefont
  {Nussbaumer}}, \bibinfo {author} {\bibfnamefont {C.}~\bibnamefont
  {Sch{\"o}nenberger}}, \ and\ \bibinfo {author} {\bibfnamefont
  {L.}~\bibnamefont {Forr\'{o}}},\ }\bibfield  {title} {\enquote {\bibinfo
  {title} {Electrochemical carbon nanotube field-effect transistor},}\
  }\href@noop {} {\bibfield  {journal} {\bibinfo  {journal} {App. Phys. Lett.}\
  }\textbf {\bibinfo {volume} {78}},\ \bibinfo {pages} {1291--1293} (\bibinfo
  {year} {2001})}\BibitemShut {NoStop}%
\bibitem [{\citenamefont {Kr{\"u}ger}\ \emph {et~al.}(2003)\citenamefont
  {Kr{\"u}ger}, \citenamefont {Widmer}, \citenamefont {Nussbaumer},
  \citenamefont {Buitelaar},\ and\ \citenamefont
  {Sch{\"o}nenberger}}]{Krueger_NewJ.Phys.5/2003}%
  \BibitemOpen
  \bibfield  {author} {\bibinfo {author} {\bibfnamefont {M.}~\bibnamefont
  {Kr{\"u}ger}}, \bibinfo {author} {\bibfnamefont {I.}~\bibnamefont {Widmer}},
  \bibinfo {author} {\bibfnamefont {T.}~\bibnamefont {Nussbaumer}}, \bibinfo
  {author} {\bibfnamefont {M.}~\bibnamefont {Buitelaar}}, \ and\ \bibinfo
  {author} {\bibfnamefont {C.}~\bibnamefont {Sch{\"o}nenberger}},\ }\bibfield
  {title} {\enquote {\bibinfo {title} {Sensitivity of single multiwalled carbon
  nanotubes to the environment},}\ }\href@noop {} {\bibfield  {journal}
  {\bibinfo  {journal} {New J. Phys.}\ }\textbf {\bibinfo {volume} {5}},\
  \bibinfo {pages} {138} (\bibinfo {year} {2003})}\BibitemShut {NoStop}%
\bibitem [{\citenamefont {Heinze}\ \emph {et~al.}(2002)\citenamefont {Heinze},
  \citenamefont {Tersoff}, \citenamefont {Martel}, \citenamefont {Derycke},
  \citenamefont {Appenzeller},\ and\ \citenamefont {Avouris}}]{Heinze02}%
  \BibitemOpen
  \bibfield  {author} {\bibinfo {author} {\bibfnamefont {S.}~\bibnamefont
  {Heinze}}, \bibinfo {author} {\bibfnamefont {J.}~\bibnamefont {Tersoff}},
  \bibinfo {author} {\bibfnamefont {R.}~\bibnamefont {Martel}}, \bibinfo
  {author} {\bibfnamefont {V.}~\bibnamefont {Derycke}}, \bibinfo {author}
  {\bibfnamefont {J.}~\bibnamefont {Appenzeller}}, \ and\ \bibinfo {author}
  {\bibfnamefont {Ph.}\ \bibnamefont {Avouris}},\ }\bibfield  {title} {\enquote
  {\bibinfo {title} {Carbon nanotubes as schottky barrier transistors},}\
  }\href@noop {} {\bibfield  {journal} {\bibinfo  {journal} {Phys. Rev. Lett.}\
  }\textbf {\bibinfo {volume} {89}},\ \bibinfo {pages} {106801} (\bibinfo
  {year} {2002})}\BibitemShut {NoStop}%
\bibitem [{\citenamefont {Foa-Torres}\ \emph {et~al.}(2014)\citenamefont
  {Foa-Torres}, \citenamefont {Roche},\ and\ \citenamefont
  {Charlier}}]{Foa-Torres_book}%
  \BibitemOpen
  \bibfield  {author} {\bibinfo {author} {\bibfnamefont {L.~E.~F.}\
  \bibnamefont {Foa-Torres}}, \bibinfo {author} {\bibfnamefont
  {S.}~\bibnamefont {Roche}}, \ and\ \bibinfo {author} {\bibfnamefont {J.-C.}\
  \bibnamefont {Charlier}},\ }\href@noop {} {\emph {\bibinfo {title}
  {{Introduction to graphene-based nanomaterials: From electronic structure to
  quantum transport}}}}\ (\bibinfo  {publisher} {CUP},\ \bibinfo {address}
  {Cambridge},\ \bibinfo {year} {2014})\BibitemShut {NoStop}%
\bibitem [{\citenamefont {Zhang}\ and\ \citenamefont
  {Levy}(1999)}]{Zhang_Eur.Phys.J.B10/1999}%
  \BibitemOpen
  \bibfield  {author} {\bibinfo {author} {\bibfnamefont {S.}~\bibnamefont
  {Zhang}}\ and\ \bibinfo {author} {\bibfnamefont {P.M.}\ \bibnamefont
  {Levy}},\ }\bibfield  {title} {\enquote {\bibinfo {title} {Models for
  magnetoresistance in tunnel junctions},}\ }\href@noop {} {\bibfield
  {journal} {\bibinfo  {journal} {Eur. Phys. J. B}\ }\textbf {\bibinfo {volume}
  {10}},\ \bibinfo {pages} {599--606} (\bibinfo {year} {1999})}\BibitemShut
  {NoStop}%
\bibitem [{\citenamefont {Mavropoulos}\ \emph {et~al.}(2004)\citenamefont
  {Mavropoulos}, \citenamefont {Papanikolaou},\ and\ \citenamefont
  {Dederichs}}]{Mavropoulos_Phys.Rev.B69/2004}%
  \BibitemOpen
  \bibfield  {author} {\bibinfo {author} {\bibfnamefont {P.}~\bibnamefont
  {Mavropoulos}}, \bibinfo {author} {\bibfnamefont {N.}~\bibnamefont
  {Papanikolaou}}, \ and\ \bibinfo {author} {\bibfnamefont {P.H.}\ \bibnamefont
  {Dederichs}},\ }\bibfield  {title} {\enquote {\bibinfo {title}
  {{Korringa-Kohn-Rostoker Green-function formalism for ballistic
  transport}},}\ }\href@noop {} {\bibfield  {journal} {\bibinfo  {journal}
  {Phys. Rev. B}\ }\textbf {\bibinfo {volume} {69}},\ \bibinfo {pages} {125104}
  (\bibinfo {year} {2004})}\BibitemShut {NoStop}%
\bibitem [{\citenamefont {Nemec}\ \emph {et~al.}(2006)\citenamefont {Nemec},
  \citenamefont {Tom\'anek},\ and\ \citenamefont
  {Cuniberti}}]{Nemec_Phys.Rev.Lett.96/2006}%
  \BibitemOpen
  \bibfield  {author} {\bibinfo {author} {\bibfnamefont {N.}~\bibnamefont
  {Nemec}}, \bibinfo {author} {\bibfnamefont {D.}~\bibnamefont {Tom\'anek}}, \
  and\ \bibinfo {author} {\bibfnamefont {G.}~\bibnamefont {Cuniberti}},\
  }\bibfield  {title} {\enquote {\bibinfo {title} {{Contact dependence of
  carrier injection in carbon nanotubes: An ab initio study}},}\ }\href@noop {}
  {\bibfield  {journal} {\bibinfo  {journal} {Phys. Rev. Lett.}\ }\textbf
  {\bibinfo {volume} {96}},\ \bibinfo {pages} {076802} (\bibinfo {year}
  {2006})}\BibitemShut {NoStop}%
\bibitem [{\citenamefont {Gijs}\ and\ \citenamefont
  {Bauer}(1997)}]{Gijs_Adv.Phys.46/1997}%
  \BibitemOpen
  \bibfield  {author} {\bibinfo {author} {\bibfnamefont {M.A.M}\ \bibnamefont
  {Gijs}}\ and\ \bibinfo {author} {\bibfnamefont {G.E.W.}\ \bibnamefont
  {Bauer}},\ }\bibfield  {title} {\enquote {\bibinfo {title} {Perpendicular
  giant magnetoresistance of magnetic multilayers},}\ }\href@noop {} {\bibfield
   {journal} {\bibinfo  {journal} {Adv. Phys.}\ }\textbf {\bibinfo {volume}
  {46}},\ \bibinfo {pages} {285--445} (\bibinfo {year} {1997})}\BibitemShut
  {NoStop}%
\bibitem [{\citenamefont {Kroemer}\ and\ \citenamefont
  {Zhu}(1982)}]{Kroemer_J.Vac.Sci.Technol.21/1982}%
  \BibitemOpen
  \bibfield  {author} {\bibinfo {author} {\bibfnamefont {H.}~\bibnamefont
  {Kroemer}}\ and\ \bibinfo {author} {\bibfnamefont {Q.-G.}\ \bibnamefont
  {Zhu}},\ }\bibfield  {title} {\enquote {\bibinfo {title} {On the interface
  connection rules for effective-mass wave functions at an abrupt
  heterojunction between two semiconductors with different effective mass},}\
  }\href@noop {} {\bibfield  {journal} {\bibinfo  {journal} {J. Vac. Sci.
  Technol.}\ }\textbf {\bibinfo {volume} {21}},\ \bibinfo {pages} {551--553}
  (\bibinfo {year} {1982})}\BibitemShut {NoStop}%
\bibitem [{\citenamefont {Zhu}\ and\ \citenamefont
  {Kroemer}(1983)}]{Zhu_Phys.Rev.B27/1983}%
  \BibitemOpen
  \bibfield  {author} {\bibinfo {author} {\bibfnamefont {Q.-G.}\ \bibnamefont
  {Zhu}}\ and\ \bibinfo {author} {\bibfnamefont {H.}~\bibnamefont {Kroemer}},\
  }\bibfield  {title} {\enquote {\bibinfo {title} {Interface connection rules
  for effective-mass wave functions at an abrupt heterojunction between two
  different semiconductors},}\ }\href@noop {} {\bibfield  {journal} {\bibinfo
  {journal} {Phys. Rev. B}\ }\textbf {\bibinfo {volume} {27}},\ \bibinfo
  {pages} {3519--3527} (\bibinfo {year} {1983})}\BibitemShut {NoStop}%
\bibitem [{\citenamefont {Harrison}(2011)}]{Harrison_J.Appl.Phys.110/2011}%
  \BibitemOpen
  \bibfield  {author} {\bibinfo {author} {\bibfnamefont {W.A.}\ \bibnamefont
  {Harrison}},\ }\bibfield  {title} {\enquote {\bibinfo {title} {Effects of
  matching conditions in effective-mass theory: Quantum wells, transmission,
  and metal-induced gap states},}\ }\href@noop {} {\bibfield  {journal}
  {\bibinfo  {journal} {J. Appl. Phys.}\ }\textbf {\bibinfo {volume} {110}},\
  \bibinfo {pages} {113715} (\bibinfo {year} {2011})}\BibitemShut {NoStop}%
\bibitem [{\citenamefont {Datta}(1997)}]{Datta_book}%
  \BibitemOpen
  \bibfield  {author} {\bibinfo {author} {\bibfnamefont {S.}~\bibnamefont
  {Datta}},\ }\href@noop {} {\emph {\bibinfo {title} {Electronic transport in
  mesoscopic systems}}}\ (\bibinfo  {publisher} {CUP},\ \bibinfo {address}
  {Cambridge},\ \bibinfo {year} {1997})\BibitemShut {NoStop}%
\bibitem [{\citenamefont {Moodera}\ \emph {et~al.}(1999)\citenamefont
  {Moodera}, \citenamefont {Nassar},\ and\ \citenamefont {Mathon}}]{Moodera99}%
  \BibitemOpen
  \bibfield  {author} {\bibinfo {author} {\bibfnamefont {J.~S.}\ \bibnamefont
  {Moodera}}, \bibinfo {author} {\bibfnamefont {J.}~\bibnamefont {Nassar}}, \
  and\ \bibinfo {author} {\bibfnamefont {G.}~\bibnamefont {Mathon}},\
  }\bibfield  {title} {\enquote {\bibinfo {title} {Spin-tunneling in
  ferromagnetic junctions},}\ }\href@noop {} {\bibfield  {journal} {\bibinfo
  {journal} {Annu. Rev. Mater. Sci.}\ }\textbf {\bibinfo {volume} {29}},\
  \bibinfo {pages} {381--432} (\bibinfo {year} {1999})}\BibitemShut {NoStop}%
\bibitem [{\citenamefont {Schmidt}\ \emph {et~al.}(2000)\citenamefont
  {Schmidt}, \citenamefont {Ferrand}, \citenamefont {Molenkamp}, \citenamefont
  {Filip},\ and\ \citenamefont {van Wees}}]{Schmidt_Phys.Rev.B62/2000}%
  \BibitemOpen
  \bibfield  {author} {\bibinfo {author} {\bibfnamefont {G.}~\bibnamefont
  {Schmidt}}, \bibinfo {author} {\bibfnamefont {D.}~\bibnamefont {Ferrand}},
  \bibinfo {author} {\bibfnamefont {L.~W.}\ \bibnamefont {Molenkamp}}, \bibinfo
  {author} {\bibfnamefont {A.~T.}\ \bibnamefont {Filip}}, \ and\ \bibinfo
  {author} {\bibfnamefont {B.~J.}\ \bibnamefont {van Wees}},\ }\bibfield
  {title} {\enquote {\bibinfo {title} {Fundamental obstacle for electrical spin
  injection from a ferromagnetic metal into a diffusive semiconductor},}\
  }\href@noop {} {\bibfield  {journal} {\bibinfo  {journal} {Phys. Rev. B}\
  }\textbf {\bibinfo {volume} {62}},\ \bibinfo {pages} {R4790--R4793} (\bibinfo
  {year} {2000})}\BibitemShut {NoStop}%
\bibitem [{\citenamefont
  {Schmidt}(2005)}]{Schmidt_J.Phys.D:Appl.Physi.38/2005}%
  \BibitemOpen
  \bibfield  {author} {\bibinfo {author} {\bibfnamefont {G.}~\bibnamefont
  {Schmidt}},\ }\bibfield  {title} {\enquote {\bibinfo {title} {Concepts for
  spin injection into semiconductors—a review},}\ }\href@noop {} {\bibfield
  {journal} {\bibinfo  {journal} {J. Phys. D: Appl. Physi.}\ }\textbf {\bibinfo
  {volume} {38}},\ \bibinfo {pages} {R107--R122} (\bibinfo {year}
  {2005})}\BibitemShut {NoStop}%
\bibitem [{\citenamefont {Rashba}(2000)}]{Rashba_Phys.Rev.B62/2000}%
  \BibitemOpen
  \bibfield  {author} {\bibinfo {author} {\bibfnamefont {E.~I.}\ \bibnamefont
  {Rashba}},\ }\bibfield  {title} {\enquote {\bibinfo {title} {{Theory of
  electrical spin injection: Tunnel contacts as a solution of the conductivity
  mismatch problem}},}\ }\href@noop {} {\bibfield  {journal} {\bibinfo
  {journal} {Phys. Rev. B}\ }\textbf {\bibinfo {volume} {62}},\ \bibinfo
  {pages} {R16267--R16270} (\bibinfo {year} {2000})}\BibitemShut {NoStop}%
\bibitem [{\citenamefont {Fert}\ and\ \citenamefont
  {Jaffres}(2001)}]{Fert_Phys.Rev.B64/2001}%
  \BibitemOpen
  \bibfield  {author} {\bibinfo {author} {\bibfnamefont {A.}~\bibnamefont
  {Fert}}\ and\ \bibinfo {author} {\bibfnamefont {H.}~\bibnamefont {Jaffres}},\
  }\bibfield  {title} {\enquote {\bibinfo {title} {Conditions for efficient
  spin injection from a ferromagnetic metal into a semiconductor},}\
  }\href@noop {} {\bibfield  {journal} {\bibinfo  {journal} {Phys. Rev. B}\
  }\textbf {\bibinfo {volume} {64}},\ \bibinfo {pages} {184420} (\bibinfo
  {year} {2001})}\BibitemShut {NoStop}%
\bibitem [{\citenamefont {Valet}\ and\ \citenamefont
  {Fert}(1993)}]{Valet_Phys.Rev.B48/1993}%
  \BibitemOpen
  \bibfield  {author} {\bibinfo {author} {\bibfnamefont {T.}~\bibnamefont
  {Valet}}\ and\ \bibinfo {author} {\bibfnamefont {A.}~\bibnamefont {Fert}},\
  }\bibfield  {title} {\enquote {\bibinfo {title} {Theory of the perpendicular
  magnetoresistance in magnetic multilayers},}\ }\href@noop {} {\bibfield
  {journal} {\bibinfo  {journal} {Phys. Rev. B}\ }\textbf {\bibinfo {volume}
  {48}},\ \bibinfo {pages} {7099--7113} (\bibinfo {year} {1993})}\BibitemShut
  {NoStop}%
\bibitem [{\citenamefont {Stone}\ and\ \citenamefont
  {Lee}(1985)}]{Stone_Phys.Rev.Lett.54/1985}%
  \BibitemOpen
  \bibfield  {author} {\bibinfo {author} {\bibfnamefont {A.~D.}\ \bibnamefont
  {Stone}}\ and\ \bibinfo {author} {\bibfnamefont {P.~A.}\ \bibnamefont
  {Lee}},\ }\bibfield  {title} {\enquote {\bibinfo {title} {Effect of inelastic
  processes on resonant tunneling in one dimension},}\ }\href@noop {}
  {\bibfield  {journal} {\bibinfo  {journal} {Phys. Rev. Lett.}\ }\textbf
  {\bibinfo {volume} {54}},\ \bibinfo {pages} {1196--1199} (\bibinfo {year}
  {1985})}\BibitemShut {NoStop}%
\bibitem [{\citenamefont {Blanter}\ and\ \citenamefont
  {B{\"u}ttiker}(2000)}]{Blanter_Phys.Rep.336/2000}%
  \BibitemOpen
  \bibfield  {author} {\bibinfo {author} {\bibfnamefont {Ya.M.}\ \bibnamefont
  {Blanter}}\ and\ \bibinfo {author} {\bibfnamefont {M.}~\bibnamefont
  {B{\"u}ttiker}},\ }\bibfield  {title} {\enquote {\bibinfo {title} {Shot noise
  in mesoscopic conductors},}\ }\href@noop {} {\bibfield  {journal} {\bibinfo
  {journal} {Phys. Rep.}\ }\textbf {\bibinfo {volume} {336}},\ \bibinfo {pages}
  {1--166} (\bibinfo {year} {2000})}\BibitemShut {NoStop}%
\bibitem [{\citenamefont {Price}(1998)}]{Price_Am.J.Phys.66/1998}%
  \BibitemOpen
  \bibfield  {author} {\bibinfo {author} {\bibfnamefont {P.~J.}\ \bibnamefont
  {Price}},\ }\bibfield  {title} {\enquote {\bibinfo {title} {Attempt frequency
  in tunneling},}\ }\href@noop {} {\bibfield  {journal} {\bibinfo  {journal}
  {Am. J. Phys.}\ }\textbf {\bibinfo {volume} {66}},\ \bibinfo {pages}
  {1119--1122} (\bibinfo {year} {1998})}\BibitemShut {NoStop}%
\bibitem [{\citenamefont {Julliere}(1975)}]{Julliere_Phys.Lett.A54/1975}%
  \BibitemOpen
  \bibfield  {author} {\bibinfo {author} {\bibfnamefont {M.}~\bibnamefont
  {Julliere}},\ }\bibfield  {title} {\enquote {\bibinfo {title} {Tunneling
  between ferromagnetic films},}\ }\href@noop {} {\bibfield  {journal}
  {\bibinfo  {journal} {Phys. Lett. A}\ }\textbf {\bibinfo {volume} {54}},\
  \bibinfo {pages} {225--226} (\bibinfo {year} {1975})}\BibitemShut {NoStop}%
\bibitem [{\citenamefont {M\"uller}\ \emph {et~al.}(2009)\citenamefont
  {M\"uller}, \citenamefont {Miao},\ and\ \citenamefont {Moodera}}]{Mueller09}%
  \BibitemOpen
  \bibfield  {author} {\bibinfo {author} {\bibfnamefont {M.}~\bibnamefont
  {M\"uller}}, \bibinfo {author} {\bibfnamefont {G.-X.}\ \bibnamefont {Miao}},
  \ and\ \bibinfo {author} {\bibfnamefont {J.S.}\ \bibnamefont {Moodera}},\
  }\bibfield  {title} {\enquote {\bibinfo {title} {Exchange splitting and
  bias-dependent transport in euo spin filter tunnel barriers},}\ }\href@noop
  {} {\bibfield  {journal} {\bibinfo  {journal} {Europhys. Lett.}\ }\textbf
  {\bibinfo {volume} {88}},\ \bibinfo {pages} {47006} (\bibinfo {year}
  {2009})}\BibitemShut {NoStop}%
\bibitem [{\citenamefont {Nagahama}\ \emph {et~al.}(2007)\citenamefont
  {Nagahama}, \citenamefont {Santos},\ and\ \citenamefont {Moodera}}]{Nag07}%
  \BibitemOpen
  \bibfield  {author} {\bibinfo {author} {\bibfnamefont {T.}~\bibnamefont
  {Nagahama}}, \bibinfo {author} {\bibfnamefont {T.S.}\ \bibnamefont {Santos}},
  \ and\ \bibinfo {author} {\bibfnamefont {J.S.}\ \bibnamefont {Moodera}},\
  }\bibfield  {title} {\enquote {\bibinfo {title} {Enhanced magnetotransport at
  high bias in quasimagnetic tunnel junctions with eus spin-filter barriers},}\
  }\href@noop {} {\bibfield  {journal} {\bibinfo  {journal} {Phys. Rev. Lett.}\
  }\textbf {\bibinfo {volume} {99}},\ \bibinfo {pages} {016602} (\bibinfo
  {year} {2007})}\BibitemShut {NoStop}%
\bibitem [{\citenamefont {G{\"o}hler}\ \emph {et~al.}(2011)\citenamefont
  {G{\"o}hler}, \citenamefont {Hamelbeck}, \citenamefont {Markus},
  \citenamefont {Kettner}, \citenamefont {Hanne}, \citenamefont {Vager},
  \citenamefont {Naaman},\ and\ \citenamefont {Zacharias}}]{Goehler11}%
  \BibitemOpen
  \bibfield  {author} {\bibinfo {author} {\bibfnamefont {B.}~\bibnamefont
  {G{\"o}hler}}, \bibinfo {author} {\bibfnamefont {V.}~\bibnamefont
  {Hamelbeck}}, \bibinfo {author} {\bibfnamefont {T.~Z.}\ \bibnamefont
  {Markus}}, \bibinfo {author} {\bibfnamefont {M.}~\bibnamefont {Kettner}},
  \bibinfo {author} {\bibfnamefont {G.~F.}\ \bibnamefont {Hanne}}, \bibinfo
  {author} {\bibfnamefont {Z.}~\bibnamefont {Vager}}, \bibinfo {author}
  {\bibfnamefont {R.}~\bibnamefont {Naaman}}, \ and\ \bibinfo {author}
  {\bibfnamefont {H.}~\bibnamefont {Zacharias}},\ }\bibfield  {title} {\enquote
  {\bibinfo {title} {Spin selectivity in electron transmission through
  self-assembled monolayers of double-stranded dna},}\ }\href@noop {}
  {\bibfield  {journal} {\bibinfo  {journal} {Science}\ }\textbf {\bibinfo
  {volume} {331}},\ \bibinfo {pages} {894--897} (\bibinfo {year}
  {2011})}\BibitemShut {NoStop}%
\bibitem [{\citenamefont {Mishra}\ \emph {et~al.}(2013)\citenamefont {Mishra},
  \citenamefont {Markus}, \citenamefont {Naaman}, \citenamefont {Kettner},
  \citenamefont {G\"{o}hler}, \citenamefont {Zacharias}, \citenamefont
  {Friedman}, \citenamefont {Sheves},\ and\ \citenamefont
  {Fontanesi}}]{Mishra13}%
  \BibitemOpen
  \bibfield  {author} {\bibinfo {author} {\bibfnamefont {D.}~\bibnamefont
  {Mishra}}, \bibinfo {author} {\bibfnamefont {T.~Z.}\ \bibnamefont {Markus}},
  \bibinfo {author} {\bibfnamefont {R.}~\bibnamefont {Naaman}}, \bibinfo
  {author} {\bibfnamefont {M.}~\bibnamefont {Kettner}}, \bibinfo {author}
  {\bibfnamefont {B.}~\bibnamefont {G\"{o}hler}}, \bibinfo {author}
  {\bibfnamefont {H.}~\bibnamefont {Zacharias}}, \bibinfo {author}
  {\bibfnamefont {N.}~\bibnamefont {Friedman}}, \bibinfo {author}
  {\bibfnamefont {M.}~\bibnamefont {Sheves}}, \ and\ \bibinfo {author}
  {\bibfnamefont {C.}~\bibnamefont {Fontanesi}},\ }\bibfield  {title} {\enquote
  {\bibinfo {title} {Spin-dependent electron transmission through
  bacteriorhodopsin embedded in purple membrane},}\ }\href@noop {} {\bibfield
  {journal} {\bibinfo  {journal} {Proc. Natl. Acad. Sci. USA}\ }\textbf
  {\bibinfo {volume} {110}},\ \bibinfo {pages} {14872--14876} (\bibinfo {year}
  {2013})}\BibitemShut {NoStop}%
\bibitem [{\citenamefont {Guo}\ and\ \citenamefont {Sun}(2014)}]{Guo14}%
  \BibitemOpen
  \bibfield  {author} {\bibinfo {author} {\bibfnamefont {A.-M.}\ \bibnamefont
  {Guo}}\ and\ \bibinfo {author} {\bibfnamefont {Q.-F.}\ \bibnamefont {Sun}},\
  }\bibfield  {title} {\enquote {\bibinfo {title} {Spin-dependent electron
  transport in protein-like single-helical molecules},}\ }\href@noop {}
  {\bibfield  {journal} {\bibinfo  {journal} {Proc. Natl. Acad. of Sci. USA}\
  }\textbf {\bibinfo {volume} {111}},\ \bibinfo {pages} {11658--11662}
  (\bibinfo {year} {2014})}\BibitemShut {NoStop}%
\bibitem [{\citenamefont {Pappas}\ \emph {et~al.}(2013)\citenamefont {Pappas},
  \citenamefont {Poulopoulos}, \citenamefont {Lewitz}, \citenamefont {Straub},
  \citenamefont {Goschew}, \citenamefont {Kapaklis}, \citenamefont {Wilhelm},
  \citenamefont {Rogalev},\ and\ \citenamefont {Fumagalli}}]{Pappas13}%
  \BibitemOpen
  \bibfield  {author} {\bibinfo {author} {\bibfnamefont {S.~D.}\ \bibnamefont
  {Pappas}}, \bibinfo {author} {\bibfnamefont {P.}~\bibnamefont {Poulopoulos}},
  \bibinfo {author} {\bibfnamefont {B}~\bibnamefont {Lewitz}}, \bibinfo
  {author} {\bibfnamefont {a.}~\bibnamefont {Straub}}, \bibinfo {author}
  {\bibfnamefont {a.}~\bibnamefont {Goschew}}, \bibinfo {author} {\bibfnamefont
  {V.}~\bibnamefont {Kapaklis}}, \bibinfo {author} {\bibfnamefont
  {F.}~\bibnamefont {Wilhelm}}, \bibinfo {author} {\bibfnamefont
  {a.}~\bibnamefont {Rogalev}}, \ and\ \bibinfo {author} {\bibfnamefont
  {P.}~\bibnamefont {Fumagalli}},\ }\bibfield  {title} {\enquote {\bibinfo
  {title} {{Direct evidence for significant spin-polarization of EuS in Co/EuS
  multilayers at room temperature.}}}\ }\href@noop {} {\bibfield  {journal}
  {\bibinfo  {journal} {Sci. Rep.}\ }\textbf {\bibinfo {volume} {3}},\ \bibinfo
  {pages} {1333} (\bibinfo {year} {2013})}\BibitemShut {NoStop}%
\bibitem [{\citenamefont {Xie}\ \emph {et~al.}(2011)\citenamefont {Xie},
  \citenamefont {Markus}, \citenamefont {Cohen}, \citenamefont {Vager},
  \citenamefont {Gutierrez},\ and\ \citenamefont {Naaman}}]{Xie11}%
  \BibitemOpen
  \bibfield  {author} {\bibinfo {author} {\bibfnamefont {Z.}~\bibnamefont
  {Xie}}, \bibinfo {author} {\bibfnamefont {T.~Z.}\ \bibnamefont {Markus}},
  \bibinfo {author} {\bibfnamefont {S.~R.}\ \bibnamefont {Cohen}}, \bibinfo
  {author} {\bibfnamefont {Z.}~\bibnamefont {Vager}}, \bibinfo {author}
  {\bibfnamefont {R.}~\bibnamefont {Gutierrez}}, \ and\ \bibinfo {author}
  {\bibfnamefont {R.}~\bibnamefont {Naaman}},\ }\bibfield  {title} {\enquote
  {\bibinfo {title} {{Spin Specific Electron Conduction through DNA
  Oligomers}},}\ }\href@noop {} {\bibfield  {journal} {\bibinfo  {journal}
  {Nano Lett.}\ }\textbf {\bibinfo {volume} {11}},\ \bibinfo {pages}
  {4652--4655} (\bibinfo {year} {2011})}\BibitemShut {NoStop}%
\end{thebibliography}

%


\end{document}